\documentclass[aps,journal=prb,twocolumn,noeprint]{revtex4-2}
\usepackage{graphicx}
\usepackage{epstopdf}
\usepackage{bmpsize}
\usepackage{braket}
\usepackage{siunitx}

\usepackage{amsmath,amsfonts,amsthm,bm}
\usepackage{bigints}
\usepackage{xcolor}
\usepackage{tikz}
\usetikzlibrary{decorations.pathreplacing}
\usepackage{placeins}
\usepackage{colortbl}
\usepackage{dsfont}
\usepackage{hhline} 
\usepackage{ragged2e}

\usepackage{tabularx}
\usepackage{array}
\usepackage[table]{xcolor}
\usepackage{amsmath}
\usepackage{makecell}

\usepackage[caption=false]{subfig}

\newcommand{\phantomsubfloat}[1]{
    {
        \captionsetup[subfigure]{labelformat=empty}
        \subfloat[][]{#1}
    }%
}

\newcommand{\ph}{\phantom{\dagger}}
\newcommand{\pd}{\phantom{\dagger}}

\usepackage{comment} 

\usepackage[bookmarks=false]{hyperref}
\hypersetup{colorlinks=true, citecolor=blue, urlcolor=blue, linkcolor=blue}

\usepackage{mathtools}

\newcommand\vertarrowbox[3][6ex]{%
  \begin{array}[t]{@{}c@{}} #2 \\
  \left\uparrow\vcenter{\hrule height #1}\right.\kern-\nulldelimiterspace\\
  \makebox[0pt]{\scriptsize#3}
  \end{array}%
}

\makeatletter
\renewcommand{\maketag@@@}[1]{\hbox{\m@th\normalsize\normalfont#1}}%
\makeatother

\def\!{\mskip-\thinmuskip}

\newcommand{\figref}[1]{\ref{#1}}
\graphicspath{{plots/}}

\begin{document}
\title{Vibrational Instabilities in Charge Transport through Molecular Nanojunctions: The Role of Anharmonic Nuclear Potentials}
\author{Martin Mäck}
\email[Corresponding author: Martin Mäck \\ Email:
]{martin.maeck@physik.uni-freiburg.de}
\affiliation{Institute of Physics, University of Freiburg, Hermann-Herder-Strasse 3, 79104 Freiburg, Germany}
\author{Michael Thoss}
\affiliation{Institute of Physics, University of Freiburg, Hermann-Herder-Strasse 3, 79104 Freiburg, Germany}
\author{Samuel L. Rudge}
\affiliation{Institute of Physics, University of Freiburg, Hermann-Herder-Strasse 3, 79104 Freiburg, Germany}

\begin{abstract}
\noindent The current-induced vibrational dynamics is a key factor determining the stability of molecular nanojunctions. Beyond conventional Joule heating, a different mechanism caused by nonconservative current-induced forces has been predicted for models with multiple vibrational modes, leading to vibrational instabilities already at low bias voltages. So far, this mechanism has only been investigated in models with harmonic nuclear potentials. Consequently, a natural question is whether this effect can also be observed in more realistic models containing anharmonic nuclear potentials, and, if so, whether it has a measurable impact on observables such as the junction dissociation probability. In this work, we apply a mixed quantum-classical approach based on electronic friction and Langevin dynamics to various anharmonic two-mode systems. By performing Langevin simulations of the vibrational dynamics, we investigate the influence of anharmonicity on instabilities arising from nonconservative forces and the corresponding dissociation dynamics of the junction, as well as steady-state observables, such as the electronic current.
\end{abstract}

\maketitle

\section{Introduction} \label{sec:Introduction}

\noindent The continuous aim to further miniaturize electronic circuits lead to the idea of building electronic devices from single molecules \cite{vonHippel1956,Aviram1974,Nitzan2001,Nitzan2003,Cuevas2010, Zimbovskaya2011, Bergfield2013, Thoss2018}. The most simple building blocks of such molecular devices, molecular nanojunctions, are constructed by attaching single molecules to two metallic leads. Beyond potential technological applications, such devices offer fascinating insight into nonequilibrium physics and can be used to explore intriguing novel physical effects that originate from the quantum mechanical nature of such systems. 

One of the most fundamental obstacles regarding the investigation of molecular nanojunctions, both in technological application and experiments, is their stability under nonequilibrium charge transport when applying a finite bias voltage. The interaction between the vibrational and electronic degrees of freedom (DOFs) within the molecule causes current-induced or Joule heating, which can lead to vibrational instabilities \cite{Segal2002,Huang2007,Hrtle2008,C1CP21161G,Tao2006,doi:10.1126/science.1146556,Ioffe2008,Montgomery2002,Pecchia2007,Schinabeck2018} and current-induced rupture of parts of the junction \cite{Erpenbeck2018,Erpenbeck2020,Ke2021, Ke2023,Li2015,Li2016,Capozzi2016,doi:10.1126/science.272.5260.385,doi:10.1021/nl801669e,PhysRevB.78.045434, Schulze2008}. 

While Joule heating may significantly influence the stability of the nanojunction at high bias voltages, there have been reports of a different mechanism of vibrational instability already at low bias voltage in systems with multiple active vibrational modes \cite{L2010, L2011,L2012,rtr4-xnny}. Generally, this mechanism can be understood in a mixed quantum-classical framework of the vibrational dynamics, in which the nuclear vibrational DOFs are treated classically and the quantum electronic DOFs are integrated out, appearing as effective electronic forces. Under further assumptions of Markovianity and weak nonadiabaticity, this yields the Markovian electronic friction and Langevin dynamics approach, in which the electronic forces explicitly separate into an adiabatic mean force, electronic friction, and a stochastic force. 

At equilibrium, these electronic forces possess certain properties. For example, at zero bias voltage, the electronic friction tensor is positive semi-definite; the second fluctuation-dissipation theorem (FDT) is satisfied; and the adiabatic mean force is conservative. However, in nonequilibrium scenarios such as at finite bias voltage, these properties are not guaranteed and, for example, electronic friction may be negative, Joule heating breaks FDT, and the adiabatic mean force can be nonconservative \cite{Dundas2009, SORBELLO1998159,PhysRevLett.114.096801,Gunst2013,Christensen2016,Lue2019,Todorov2010}. It is exactly this last property which provides the basis for the multi-mode mechanism of vibrational instability under investigation in this work. Specifically, if two or more vibrational frequencies in a multi-mode system are degenerate, the nuclear trajectory follows a path with fixed rotational direction and, hence, gains energy in each cycle from the nonconservative force field \cite{Dundas2009}, eventually leading to an instability.

Although this mechanism is well understood for degenerate vibrational modes, there have been differing reports about whether it survives in nondegenerate multi-mode systems. For example, in Refs.~\cite{L2010,L2012}, the authors used an eigenmode analysis to predict that the the Berry force \cite{L2010,L2012,PhysRevLett.107.036804,Todorov2014}, which corresponds to the antisymmetric component of the electronic friction tensor, can pull the vibrational trajectories into elliptical shapes and generating a vibrational instability, even for nondegenerate vibrational modes. In contrast, we recently used a fully coordinate-dependent Markovian Langevin equation to simulate the dynamics of the same system as in Ref.~\cite{L2012}, and only observed instabilities for truly degenerate vibrational modes~\cite{rtr4-xnny}. For nondegenerate modes, the dynamics remained stable and the Berry force had a negligible impact on the dynamics.


Furthermore, while the mechanism has consistently been reported for degenerate vibrational modes, all investigations have so far only treated harmonic vibrational modes. However, realistic molecular systems are generally at least slightly anharmonic. Therefore, given that the mechanism is highly sensitive to frequency detuning, a natural question is whether it is also highly sensitive to anharmonicity in the nuclear potentials. Moreover, if one can observe this mechanism of vibrational instability in anharmonic systems, such as dissociative potentials, would it then have a measurable impact on bond rupture and the corresponding dissociation dynamics? These questions are highly relevant even for systems such as those in Ref.~\cite{PhysRevB.107.085419}, in which the vibrational frequencies may indeed be close to degenerate, but are unlikely to be purely harmonic and may give an important hint as to why experimental evidence of this mechanism is scarce \cite{Sabater2015}. 

Motivated by these questions, in this work, we explore the vibrational dynamics of various degenerate and nondegenerate anharmonic multi-mode systems. We use the same theoretical techniques as previous work, treating the vibrational dynamics via a Markovian Langevin equation \cite{HeadGordon1995,Bode2012,L2012, Dou2016, Dou2016_2,Maurer2016, Dou2017, Dou2017_1, Dou2018, Chen2018, Chen2019, Preston2021, PhysRevB.83.115420,10.1063/5.0019178,Rudge2023,Rudge2024,10.1063/5.0222076} and calculating the electronic forces via NEGFs \cite{Dou2017,Dou2016, Dou2016_2, Preston2020,Preston2021,Preston2022}, which forms the combined (NEGF-LD) approach. Specifically, we explore a realistic two-level, two-mode dissociatve model containing Morse potentials instead of harmonic oscillators, as well as a simpler quartic model that allows for rigorous investigation of the anharmonic effects. Via dissociation rates and the steady-state vibrational energy, we show that the mechanism of vibrational instability arising from nonconservative electronic forces disappears for even slightly anharmonic potentials, similar to previous results on vibrational frequency detuning in models containing two harmonic modes \cite{rtr4-xnny}.


The paper is structured as follows. In Sec.~\ref{sec: Model}, we introduce a general two-level, two-mode model of a molecule interacting with metallic leads. In Sec.~\ref{sec: Nonequilibrium Transport Theory}, we give a short overview of the nonequilibrium transport theory in this work, first demonstrating how to obtain electronic forces from NEGFs and then describing the vibrational dynamics with a Markovian Langevin equation. In Sec.~\ref{sec: Results}, we apply the approach to various model systems for which we simulate the vibrational dynamics, thereby analyzing the effect of anharmonic nuclear potentials on their stability and steady state observables. 

Since we have to differentiate between classically and quantum mechanically treated coordinates and momenta, we will explicitly denote the vectors of position and momentum operators as $\hat{\bm{x}}$ and $\hat{\mathbf{p}}$, while their classical counterparts will be denoted by $\bm{x}$ and $\mathbf{p}$. Moreover, in this work, we use units where $e=\hbar=1$.

\section{Model} \label{sec: Model}
In this section, we introduce the general model of a molecular nanojunction, and specify it for the particular models considered within this work, thereby expanding the model considered in Ref.~\cite{rtr4-xnny} to anharmonic potentials. 
The total Hamiltonian of the setup is 
\begin{align}
    H = \: & H_{\text{mol}} + H_{\text{leads}} + H_{\text{mol-leads}},
\end{align}
where $H_{\text{mol}}$ is the molecular Hamiltonian, $H_{\text{leads}}$ is the Hamiltonian of the leads, and $H_{\text{mol-leads}}$ describes the interaction between the two.

In our model, the molecular Hamiltonian consists of a vibrational component, $H^{\text{vib}}_\text{mol}$, and a term accounting for both the purely electronic contribution and the electronic–vibrational interactions, $H^{\text{el}}_\text{mol}$:
\begin{align} 
    H_\text{mol} = \: & \underbrace{\sum_{mn} h_{mn}(\hat{\bm{x}})d^\dagger_{m} d^{\pd}_{n}}_{= H^{\text{el}}_\text{mol}} + \underbrace{\sum_{i}\left(U_{\mathrm{u},i}\left(\hat{x}_{i}\right) + \frac{\hat{p}_{i}^{2}}{2m_i}\right)}_{= H^{\text{vib}}_\text{mol}}. \label{eq: molecular Hamiltonian}
\end{align}
The operators $d^{\dag}_{m}$ and $d^{\pd}_{m}$ create and annihilate an electron at the $m$th energy level with energy $h_{mm}(\hat{\bm{x}})$, respectively. The off-diagonal elements  $h_{mn}(\hat{\bm{x}})$ refer to the electron hopping between the electronic levels. 
Moreover, $U_{\mathrm{u},i}(\hat x_i)$ denotes the potential of the uncharged state of the molecule along vibrational mode $i$. 
The $i$th vibrational DOF has coordinate $\hat{x}_{i}$, and momentum $\hat{p}_{i}$. Since we will consider exclusively models with two vibrational modes we collect them into vectors $\hat{\bm{x}} = \left(\hat{x}_{1},\hat{x}_{2}\right)$ and $\hat{\mathbf{p}} = \left(\hat{p}_{1},\hat{p}_{2}\right)$, respectively. 

The energies of the electronic DOFs are contained in the matrix of single-particle energies and interactions, $h(\hat{\bm{x}})$, which in this work will always have the form
\begin{align} \label{eq:hamiltonian_2x2}
    h(\hat{\bm{x}}) = \: & \left(\begin{array}{c c}
      U^{(1)}_\mathrm{c}(\hat x_2)-U_{\mathrm{u},2}(\hat x_2)  & t(\hat x_1) \\
       t(\hat x_1) &   U^{(2)}_\mathrm{c}(\hat x_2)-U_{\mathrm{u},2}(\hat x_2)
    \end{array}\right).
\end{align}
Here, $ U^{(1)}_\mathrm{c}(\hat x_2)$ and $U^{(2)}_\mathrm{c}(\hat x_2)$ denote the potentials of the charged states (1) and (2).  The off-diagonal terms, $t(\hat x_1)$, describe vibrationally assisted hopping between the electronic levels. 
The specific form of the introduced potentials used in this work will be given in Sec.~\ref{sec: Results}. Note that, in general, both vibrational modes may couple diagonally and off-diagonally to both electronic levels. In principle, such systems can also be explored the NEGF-LD framework introduced in Sec.~\ref{sec: Nonequilibrium Transport Theory}. In this work, however, we have chosen the explicit form of Eq.\eqref{eq:hamiltonian_2x2} as it represents the simplest anharmonic extension of previous investigations into instabilities arising from current-induced forces in multi-mode systems \cite{rtr4-xnny,L2010,L2011,L2012,Lue2019}.

To this end, throughout this work we further restrict the anharmonicity to mode $2$, such that the potential of mode $1$ has the form 
\begin{equation}
    U_{\mathrm{u},1}(\hat x_1) =  \frac{1}{2}m_1\omega_1^2 \hat x_1^2.
\end{equation}
Moreover, the hopping between the electronic levels is assumed to always couple linearly to coordinate $\hat{x}_1$, such that
\begin{equation}
    t( \hat x_1) = t_0+\lambda_1 \hat x_1.
\end{equation}

Furthermore, to compare our results to  previous work \cite{rtr4-xnny}, we will also consider the corresponding harmonic version of Eq.\eqref{eq:hamiltonian_2x2},
which we obtain by performing a Taylor expansion of $U_{\mathrm{u},2}(x_2)$, $ U^{(1)}_\mathrm{c}(\hat x_2)$, and $U^{(2)}_\mathrm{c}(\hat x_2)$ up to second order around their respective minima. For example, the corresponding harmonic potential of $U_{\mathrm{u},2}(\hat x_2)$ is 
 \begin{equation} \label{eq: harm_approx_Ue2}
 \begin{aligned}
     U^{\mathrm{harm}}_{\mathrm{u},2}(\hat x_2) 
     = &
     \left. U^{\mathrm{}}_{\mathrm{u},2}\left(\hat x_2\right)\right|_{\hat{x}^{0}_2} 
     + 
     \left. \frac{\partial}{\partial \hat x_2} U^{\mathrm{}}_{\mathrm{u},2}\left(\hat x_2\right)\right| _{\hat{x}^{0}_2} \left(\hat x_2-\hat x^0_2 \right)
     \\
     &+
     \frac{1}{2}\left.  \frac{\partial^2}{\partial \hat x_2^2}U^{\mathrm{}}_{\mathrm{u},2} \left(\hat x_2\right) \right|_{\hat{x}^{0}_2} \left(\hat x_2-\hat x^0_2 \right)^2.
 \end{aligned}
 \end{equation}
Throughout this work, we choose the potential $U^{\mathrm{harm}}_{\mathrm{u},2}(\hat x_2)$ such that the minimum of the potential is at $\hat x^0_2=0$. 
Using a similar procedure, we obtain the corresponding harmonic versions of $ U^{(1)}_\mathrm{c}(\hat x_2)$, and $U^{(2)}_\mathrm{c}(\hat x_2)$. 

Finally, the harmonic analogue of Eq.\eqref{eq:hamiltonian_2x2} is given by
\begin{align} \label{eq: hamiltonian_2x2_harm}
    h^{\mathrm{harm}}(\hat{\bm{x}}) = \: & \left(\begin{array}{c c}
     \varepsilon_0 +\lambda_2 \hat x_2  & t_0+\lambda_1\hat x_1 \\
        t_0+\lambda_1\hat x_1  &   -\varepsilon_0 -\lambda_2 \hat x_2
    \end{array}\right),
\end{align}
where $\lambda_i=\tilde\lambda_{i}\sqrt{m_{i} \omega_{i}}$, and the potentials of both coordinates in the uncharged state are now harmonic,
\begin{equation}\label{eq: nuclear_potential_harm}
   U^{\mathrm{harm}}_{\mathrm{u},x_i}\left(\hat{x}_{i}\right)= \frac{1}{2} m_i \omega^2_i \hat x_i^2. 
\end{equation}
The parameters $ \tilde \lambda_i, \omega_i, \epsilon_0 $ and $t_0$ are determined by 
the Taylor expansion of $U_{\mathrm{u},i}(\hat x_i)$, $ U^{(1)}_\mathrm{c}(\hat x_2)$, and $U^{(2)}_\mathrm{c}(\hat x_2)$.

The left $(\mathrm {L})$ and right $(\mathrm {R})$ leads are modeled as reservoirs of noninteracting electrons, 
\begin{equation}
    H_{\text{leads}} =  \sum_{\alpha \in \{\mathrm {L},\mathrm {R}\}} \sum_{k } \varepsilon^{\ph}_{k \alpha} c^{\dagger}_{k \alpha}c^{\ph}_{k \alpha}.
\end{equation} 
The energy of state $k$ in lead $\alpha$ is given by $\varepsilon_{k \alpha}$, while $c^{\dagger}_{k \alpha} $ and $c_{k \alpha}$ denote the corresponding creation and annihilation operators, respectively. Both leads are held at local equilibrium, such that they have a well-defined chemical potential, $\mu_{\alpha}$, and temperature, $T$. By applying a bias voltage to the junction, $\Phi = \mu_\mathrm{L} - \mu_\mathrm{R}$ with $\mu_{\mathrm{L}} = -\mu_{\mathrm {R}} = e\Phi/2$, the system is driven out of equilibrium. 

The interaction between the molecule and the leads is given by
\begin{align}
    H_{\text{mol-leads}} = \: & \sum_{k,\alpha}\sum_{m} V_{k\alpha,m}\left(c_{k {\alpha}}^{\dagger}d^{\ph}_m +d_{m}^{\dagger}c^{\ph}_{k {\alpha}}\right),
\end{align}
where $V_{k\alpha,m}$ describes the coupling strength between state $m$ in the molecule and state $k$ in lead $\alpha$. For noninteracting reservoirs and a molecule-lead coupling linear in the respective creation and annihilation operators, the influence of the leads on the molecular dynamics is completely described by two-time correlation functions, which are in turn characterized in terms of the spectral density of each lead,
\begin{align}
    \Gamma_{\alpha,m m'} (\epsilon) = \: & 2 \pi \sum_{k} V^{\ph}_{k\alpha,m}V^{\ph}_{k\alpha,m'} \delta(\epsilon - \varepsilon_{k \alpha}).
\end{align}
We will work in the wideband limit, such that the spectral density is a constant: $\Gamma_{\alpha,m m'} (\epsilon) = \Gamma_{\alpha,m m'}$. Moreover, we exclusively consider models where only level $(1)$ couples to the left lead and level $(2)$ to the right lead, such that $\Gamma_{\mathrm {L},22} = \Gamma_{\mathrm{R},11} = \Gamma_{\alpha,21} = \Gamma_{\alpha,12} = 0$. Moreover, we assume that the remaining coupling strengths are the same: $\Gamma_{\mathrm {L},11} = \Gamma_{\mathrm {R},22} = \Gamma$, and will refer to $\Gamma$ as the molecule-lead coupling strength.

\section{Nonequilibrium Transport Theory} \label{sec: Nonequilibrium Transport Theory}

\noindent In this section, we briefly introduce and discuss the approaches we use to investigate the influence of anharmonic potentials on vibrational instabilities in molecular nanojunctions. More detailed overviews can be found in Refs.~\cite{rtr4-xnny,Preston2020,Preston2021,Preston2022,Chen2018,Chen2019,Rudge2023,L2012}.

First, in Sec.~\ref{subsec: Electronic Friction and Langevin Dynamics}, we discuss the approximations leading to a Markovian mixed quantum-classical Langevin equation from the fully quantum vibrational dynamics, and provide expressions for the resulting electronic forces in terms of NEGFs. Then, in Sec.~\ref{subsec: Langevin simulations and Observables}, we discuss the numerical details of solving the Langevin equation and how we calculate dissociation probabilities and expectation values of observables in the steady-state within this approach. 

\subsection{NEGF and Langevin Dynamics (NEGF-LD)} \label{subsec: Electronic Friction and Langevin Dynamics}

In this section, we give a short overview of the Markovian electronic friction and Langevin equation approach, before connecting it to the Keldysh NEGFs formalism. 


The time evolution of the fully quantum mechanically treated vibrational DOFs can be expressed through the Feynman-Vernon influence functional, which incorporates the effect of the electronic DOFs in the molecule and leads via an effective action. As shown in Refs.~\cite{L2012,Chen2018}, a classical equation of motion for the vibrational DOFs is obtained by transforming to Wigner coordinates and expanding to second order in the quantum difference to the classical path. In this limit, the quantum electronic DOFs are  integrated out, and they influence the vibrational dynamics as effective electronic forces. 

Under the further assumption of weak nonadiabaticity in the form of a timescale separation between fast quantum electronic DOFs and slow classical vibrational DOFs, these forces can be further expanded, yielding a Markovian Langevin equation:
\begin{align} \label{eq: langevin_equation}
    m_i\ddot{x}_{i} = \: & - \frac{\partial U_{\text{u,}i}}{\partial x_{i}} + F^{\text{ad}}_i(\bm{x}) - \sum_{j} \gamma_{ij}(\bm{x}) \dot{x}_{j} +  f_i(t).
\end{align}
The first term on the righthand side of Eq.\eqref{eq: langevin_equation} describes the force originating solely from the vibrational potentials, while $F^{\text{ad}}_i(\bm{x})$ is the adiabatic contribution to the mean electronic force, which is calculated at a frozen vibrational frame $\boldsymbol{x}$. It can be shown that $F^{\text{ad}}_i(\bm{x})$ is conservative for single-mode systems or multi-mode systems in equilibrium. In contrast, for multi-mode systems out of equilibrium this is not guaranteed.

Next, the third term describes the influence of the electronic friction, which arises as a first-order nonadiabatic correction to the adiabatic contribution to the mean electronic force. Finally, $f_i(t)$ is a Gaussian random force with white noise,
\begin{align} \label{eq:white_noise}
    \langle f_{i}(t)f_{j}(t) \rangle = \: & D_{ij}(\bm{x}) \delta(t-t').
\end{align}
In equilibrium, the electronic friction tensor is guaranteed to be positive semi-definite and is related to the correlation function of the stochastic force via the classical fluctuation-dissipation theorem,
\begin{align}
    D_{ij}(\bm{x}) = k_{B}T\gamma_{ij}(\bm{x}).
\end{align}
Similarly to $F^{\text{ad}}_i(\bm{x})$, however, these properties are not guaranteed at finite bias voltage, which leads, for example, to effects such as Joule heating. 

The electronic forces can be calculated in the Keldysh NEGFs framework via 
\begin{equation}\label{eq: adiabatic_force}
    F^{\text{ad}}_i(\bm{x}) = i\int^\infty_{-\infty} \frac{d\varepsilon}{2\pi} \text{Tr} \left\{ \frac{\partial h (\bm{x})}{\partial x_i} \tilde G_{(0)}^<(\varepsilon, \bm{x})\right\},
    \end{equation}
\begin{equation}
    \gamma_{ij}(\bm{x}) = -i\int^\infty_{-\infty} \frac{d\varepsilon}{2\pi} \text{Tr} \left\{ \frac{\partial h(\bm{x})}{\partial x_i} \tilde G_{(1),x_j}^<(\varepsilon, \bm{x})\right\},
    \label{friction_tensor}
    \end{equation}
and \cite{Preston2022}
\begin{equation}\label{eq: corrfunc}
   D_{ij}(\bm{x}) = \int^\infty_{-\infty} \frac{d\varepsilon}{2\pi} \text{Tr} \left\{ \frac{\partial h(\bm{x})}{\partial x_i} \tilde G_{(0)}^>(\varepsilon, \bm{x}) \frac{\partial h(\bm{x})}{\partial x_j} \tilde G_{(0)}^<(\varepsilon, \bm{x})\right\}.
\end{equation}   

The adiabatic lesser/greater Green's function in Eq.\eqref{eq: adiabatic_force} and Eq.\eqref{eq: corrfunc}, $\tilde G_{(0)}^{</>}$, is given by  
\begin{equation}
\tilde G_{(0)}^{</>} = \tilde G_{(0)}^R \tilde \Sigma^{</>} \tilde G_{(0)}^A,
\end{equation}
where the retarded/advanced Green's functions take the standard form,
\begin{equation}
\tilde{G}_{(0)}^{R/A} = \Big(\varepsilon \hat{I}-h-\tilde{\Sigma}^{R/A}\Big)^{-1}.
\label{GRA}
\end{equation}
By taking the wideband limit for the leads, the self-energies for lead $\alpha$ take the form 
\begin{equation}
\tilde{\Sigma}^R_{\alpha} = -\frac{i}{2}\Gamma_{\alpha},
\;\;\;\;\;\;
\tilde{\Sigma}^A_{\alpha} = \frac{i}{2}\Gamma_{\alpha},
\label{sigmaAR}
\end{equation}
\begin{equation}
\tilde{\Sigma}^<_{\alpha}(\varepsilon) = if_\alpha(\varepsilon)\Gamma_{\alpha},
\;\;\;\;\;\;
\tilde{\Sigma}^>_{\alpha}(\varepsilon) = -i[1-f_\alpha(\varepsilon)]\Gamma_{\alpha},
\label{sigma_lesser}
\end{equation}
with the Fermi-Dirac function given by
\begin{equation}
    f_{\alpha}(\varepsilon)= \frac{1}{1+e^{ \left(\varepsilon -\mu_\alpha \right)/k_{\text{B}}T}}.
\end{equation}
The total self-energy, which is a sum over the self-energy contributions from each lead, is written without a lead subscript, $\tilde{\Sigma}^R=\sum_\alpha \tilde{\Sigma}^R_{\alpha}$.

Moreover, the first nonadiabatic correction to the adiabatic lesser and retarded Green's function,$\tilde G_{(1),x_j}^<$ and $\tilde{G}_{(1),x_j}^{R}$, in Eq.\eqref{friction_tensor}  are given by \cite{Preston2021}
\begin{align}
\tilde{G}_{(1),x_j}^{<} = \: & \frac{1}{2i}\tilde{G}_{(0)}^{R}\Bigg(\tilde \Sigma^< \tilde G^A_{(0)}\left[\tilde G^A_{(0)},\frac{\partial h}{\partial x_j}\right]_- \nonumber \\
& + \frac{\partial h}{\partial x_j}\tilde{G}_{(0)}^{R}\frac{\partial\tilde{\Sigma}^{<}}{\partial\varepsilon}+\tilde{G}_{(0)}^{<}\frac{\partial h}{\partial x_j} + \text{h.c.}\Bigg)\tilde{G}_{(0)}^{A},
  \label{G1L} \\
\tilde{G}_{(1),x_j}^{R} = \: & \frac{1}{2i}\tilde{G}_{(0)}^{R}\left[\tilde{G}_{(0)}^{R},\frac{\partial h}{\partial x_j}\right]_{-}\tilde{G}_{(0)}^{R}.
\label{G1AR}
\end{align}

The electronic friction approach is justified for systems with a clear timescale separation between the electronic and vibrational DOFs. While the electronic timescales are mostly determined by the molecule-lead coupling, $\Gamma$, the vibrational timescales are determined by the effective harmonic frequencies of the vibrational DoFs, $\omega_{i}$. Since all results in our investigation are obtained in the limit $\Gamma \gg \omega_{i}$, this condition is fulfilled and we can expect the electronic friction approach to be valid. 

\subsection{Langevin Simulations and Observables} \label{subsec: Langevin simulations and Observables}

All observables of interest are obtained by simulating the vibrational dynamics of our model systems by solving the Langevin equation in Eq.\eqref{eq: langevin_equation}. In this work, we use the ABOBA algorithm, an efficient method for integrating stochastic differential equations.
The approach splits the generator of the Langevin equation into three components (A,B,O) using a Trotter decomposition and applies them in a specific sequence. For a more comprehensive outline of the method, see Ref.~\cite{e19120647}. In the context of charge transport in molecular nanojunctions this algorithm has previously also been applied in Refs.~\cite{Rudge2024,rtr4-xnny}.

Since the Langevin equation describes a stochastic process, we need to average over many trajectories
to calculate meaningful averages of the observables of interest. Assuming that the molecule is initially in its vibrational ground state, we sample $N_{\text{traj}}$ initial conditions from the Wigner distribution of the ground state of a two-dimensional harmonic oscillator,
\begin{equation}
    \rho_{\text{W}}(\bm{x},\bm{p})=\prod_{i}  \frac{1}{\pi} e^{-m_i\omega_i\left(x_{i} - x^{0}_{i} \right)^2-\frac{1}{m_i \omega_i}p^{2}_{i}}.
\end{equation}
For the anharmonic potentials considered here, the frequency $\omega_i$ is obtained by the harmonic approximation to the uncharged potentials $U_{\mathrm{u},2}(x_2)$ at the respective minima of their potential wells, $x^{0}_{2}$, as given in Eq.\eqref{eq: harm_approx_Ue2}. Starting from this sampling procedure, the trajectories are then propagated in time using the ABOBA algorithm mentioned above. 

For the dissociative models in this paper, we define a certain threshold position $\bm{x}_{\mathrm{diss}}$. Each time a trajectory exceeds this position, the molecular junction is assumed to be dissociated. Based on this definition of $\bm{x}_{\mathrm{diss}}$, we calculate time-dependent dissociation probabilities via an ensemble average over the trajectories,
\begin{equation}
    P_{\mathrm{diss}}(t) =\frac{1}{N_{\mathrm{traj}}} \sum_{\substack{n \in N_{\text{traj}} }} \Theta \left( \bm{x}_n(t)-\bm{x}_{\text{diss}}\right)\cdot 100.
\end{equation}
In contrast, in the non-dissociative models, we use the average vibrational energy of each mode as a measure of instability, $\langle E_{i} \rangle (t)$. At time $t$, this is obtained by 
\begin{align} \label{observable}
    \langle E_{i} \rangle (t) = \: & \frac{1}{N_{\mathrm{traj}}}  \sum_{j = 1}^{N_{\mathrm{traj}}} E_{i}(\bm{x}_{j}(t),\bm{p}_{j}(t)) \\
    E_i(\bm{x}_{j}(t),\bm{p}_{j}(t)) = \: & U_{\mathrm{u},i}(x_{i,j}(t)) + \frac{p_{i,j}(t)^2}{2m_i},
\end{align}
where $\left(x_{i,j}(t),p_{i,j}(t)\right)$ refers to the phase-space position of vibrational mode $i$ in trajectory $j$ at time $t$. Note that this expression only considers the potential of the unperturbed nuclear potentials and does not include the energy originating from the electronic-vibrational interaction.

Furthermore, we also investigate steady-state quantities, which we obtain by propagating $N_{\text{traj}}$ initial conditions to a  time $t_{\text{ss}}$ in which both $\langle E_{i} \rangle (t)$ are constant. Upon reaching the steady state, we propagate the trajectories further and sample $N_{\text{sample}}$ equally spaced points, yielding a total of $N_{\text{tot}} = N_{\text{traj}} \times N_{\text{sample}}$ points from which the expectation values of observables are computed. Following this procedure, we evaluate steady-state expectation values of the vibrational kinetic energies as 
\begin{equation}\label{observable}
    \langle E_{i} \rangle^{\text{ss}}  = \frac{1}{N_{\text{tot.}}} \sum_{j = 1}^{N_{\text{tot.}}} E_i\left(\bm{x}_{j},\bm{p}_{j}\right).
\end{equation}


Similarly, the steady-state expectation values of an electronic observable $O$ can be obtained by replacing the kinetic energy variable with the corresponding phase-space-dependent quantum expectation value, $\langle O \rangle_{\text{el}}(\bm{x}_{j},\bm{p}_{j})$:
\begin{equation}\label{observable}
    \langle O \rangle^{\text{ss}}  = \frac{1}{N_{\text{tot.}}} \sum_{j = 1}^{N_{\text{tot.}}} \langle O \rangle_{\text{el}}(\bm{x}_{j},\bm{p}_{j}).
\end{equation}
We are specifically interested in the electric current from lead $\alpha$, which in the near-adiabatic limit has the form 
\begin{equation}
    \langle I_\alpha \rangle_\text{el} (\bm{x},\bm{p}) \approx \langle I^\text{ad}_\alpha  \rangle_\text{el} (\bm{x}) + \langle I^\text{na}_\alpha \rangle_\text{el}(\bm{x},\bm{p}),
\end{equation}
Here, $\langle I_\alpha \rangle_\text{el} (\bm{x},\bm{p})$ has been separated into an adiabatic contribution, $\langle I^\text{ad}_\alpha (\bm{x})\rangle$, which depends instantaneously on the molecular coordinates, and a first-order nonadiabatic correction, $\langle I^\text{na}_\alpha(\bm{x},\bm{p})\rangle$, which additionally accounts for the non-zero momenta of the molecule. The adiabatic term is given by \cite{PhysRevLett.68.2512}
\begin{align}
\langle I_\alpha^\text{ad} \rangle_{\text{el}} (\bm{x}) = \: & \frac{1}{\pi}\int_{-\infty}^{\infty}d\varepsilon \: \text{Re}\left(\text{Tr}\left\{ \tilde{G}_{(0)}^{<}\tilde{\Sigma}_{\alpha}^{A}+\tilde{G}_{(0)}^{R}\tilde{\Sigma}_{\alpha}^{<}\right\}\right), \label{eq: instantaneous adiabatic current}
\end{align}
while the nonadiabatic correction is \cite{Preston2020,PhysRevLett.107.036804}
\begin{align}
\langle I_\alpha^\text{na} \rangle_{\text{el}} (\bm{x},\bm{p}) = & \sum_j  I_{\alpha,j}^{\mathrm{na}}(\bm{x}) \frac{p_j}{m_j}
\\
I_{\alpha,j}^{\mathrm{na}}(\bm{x}) = & \int_{-\infty}^{\infty}\frac{d\varepsilon }{\pi}\text{Re}\left(\text{Tr}\left\{ \tilde{G}_{(1),x_j}^{<}\tilde{\Sigma}_{\alpha}^{A}+ \tilde{G}_{(1),x_j}^{R}\tilde{\Sigma}_{\alpha}^{<}\right\}\right). \label{eq: instantaneous nonadiabatic current}
\end{align}
Note that these quantum expectation values have implicitly been calculated in the instantaneous electronic steady-state.


   

\section{Results}\label{sec: Results}

In this section, we will explore the influence of anharmonic nuclear potentials on nanojunction stability by performing Langevin simulations of the vibrational dynamics. First, we will investigate the vibrational dynamics and the dissociation dynamics of the two-level two mode-system  introduced in Sec.~\ref{sec: Model} with Morse potentials for the uncharged and charged states. By expanding the Morse potentials up to second order, we directly compare the vibrational dynamics of the anharmonic model with the corresponding harmonic case. Next, in order to obtain a quantitative understanding of the effect of anharmonicities on vibrational instability, we consider quartic potentials, which can be easily tuned. Note that our aim is to investigate effects of anharmonic potentials on the stability and dissociation dynamics of nanojunctions in general, and not of a specific molecule.

\subsection{ Vibrational Dynamics in a Dissociative Model} \label{subsec: diss_dynamics}

\begin{figure} 
     \centering
     \includegraphics[width=\columnwidth, trim=10 7 10 10, clip]{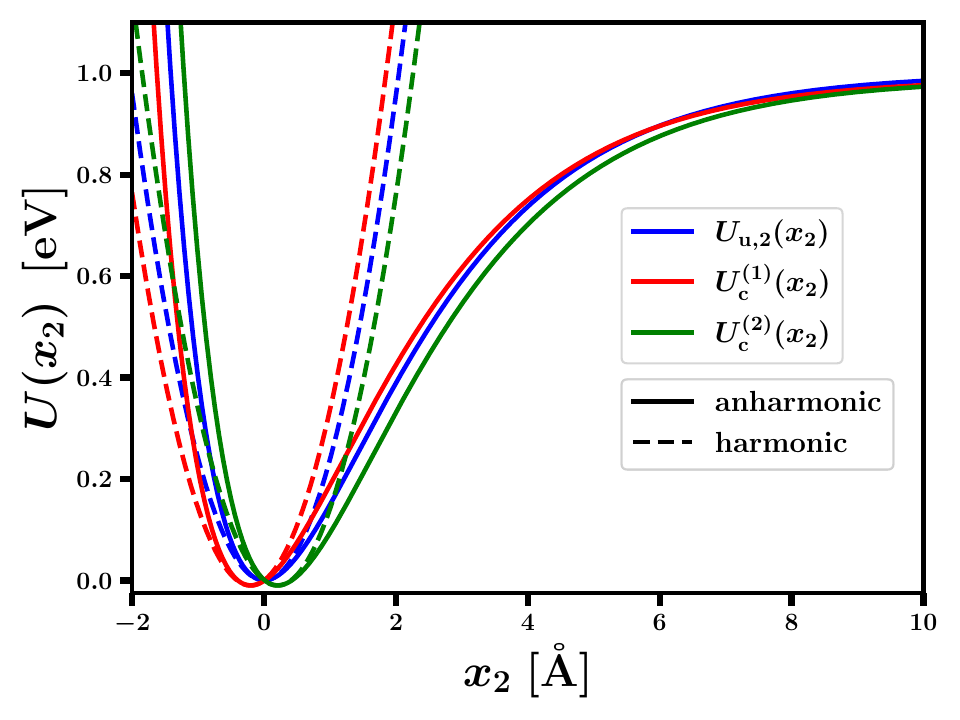}
     \caption{Potentials $ U_{\mathrm{u},2}(x_2),~U^{(1)}_{\text{c}}(x_2),$ and $U^{(2)}_{\text{c}}(x_2)$ of the uncharged state and the charged states (1) and (2), as well as their harmonic approximations. The parameters are the same as in Set (1) in Tbl.~\ref{tab: paras_table}, with $\omega_2=20~\mathrm{meV}$. }
     \label{fig: potentials_parameter_set_1}
 \end{figure}

We start by specifying the parameters and potentials of the general model introduced in Sec.~(\ref{sec: Model}). The two electronic levels (1) and (2) describe two sites of a molecular bridge, such as in a diatomic molecule, and each couple only to the closest lead. In this scenario, the vibrational coordinate $x_2$ describes the center of mass motion of the entire molecule, while the vibrational coordinate $x_1$ describes the bond-stretching motion between both parts of the molecule. The raw nuclear potential of vibrational coordinate $x_2$ is given by 
\begin{align}
    U_{\mathrm{u},2}(x_2) = \: & D^{\pd}_{\mathrm{u}}\left( 1-e^{-a_{\mathrm{u}} \left({x_{2}-x^{\mathrm{0}}_{2}}\right)}\right)^2 \\
\end{align}
while the potential of charged state $m$ is 
\begin{align}
    U^{(m)}_{\text{c}}(x_2) = \: & D_{\text{c}}^{(m)}\left( 1-e^{-a^{(1)}_{\text{c}} \left({x_{2}-x^{(m)}_{\text{c}}}\right)}\right)^2+V^{(m)}_{\text{c}}.
\end{align}
Here, $x^{({m})}_{\text{c}}$ describes the shift in the equilibrium geometry of the molecule upon state $(m)$ becoming occupied. For large displacements from the equilibrium geometry, the molecule can dissociate along the $x_2$ coordinate, with the dissociation threshold taken as $x^{\mathrm{diss}}_2=10~\mathrm{\mathring{A}}$. These potentials are illustrated in Fig.~\figref{fig: potentials_parameter_set_1}. The functional dependence of the potential $ U_{\mathrm{u},1}(x_1) $ and the hopping between the electronic levels has already been given in Sec.~\ref{sec: Model}. 

The parameters of  $ U_{\mathrm{u},2}(x_2),~U^{(1)}_{\text{c}}(x_2),$ and $U^{(2)}_{\text{c}}(x_2)$ are determined by their respective harmonic approximations, as outlined in Sec.~\ref{sec: Model}, and are given by
\begin{align}
    a_\mathrm{u} = \: & \omega_2\sqrt{ \frac{m_2}{2 D_{\mathrm{u}}}}, \qquad ~ a^{(m)}_\text{c}= \omega_2\sqrt{\frac{m_2}{2 D^{(m)}_{\text{c}}} } , \\
     x^\mathrm{0}_2 = \: & 0, \qquad \qquad \:\:\:\:\:\:\:\: ~ x^{(m)}_{\text{c}} = -\tilde \lambda_m / \sqrt{ m_2 \omega^3_2}
 \end{align}
and
\begin{equation}
    V^{(m)}_{\text{c}} = \varepsilon_0- (\tilde\lambda_m )^2/(2\omega_2).
\end{equation}
The parameters $\tilde \lambda_m, \omega_i,\varepsilon_0 $, and $t_0$ are chosen such that the harmonic approximation to the Morse potentials yields a harmonic system with similar parameters as in Refs.~\cite{L2012,rtr4-xnny}, where the mechanism for vibrational instability has been thoroughly studied. The explicit parameters used given in Tbl.~\ref{tab: paras_table}. Note that we use two parameter sets, which differ only in the electronic-vibrational couplings, $\tilde \lambda_i$. 

\begin{table}[t]
\centering
\renewcommand{\arraystretch}{1.5} 

\vspace{-1.5ex} 

\begin{tabular}{c c c}

\textbf{Parameters} & \textbf{Set (1)} & \textbf{Set (2)} \\
\hline
$\omega_{1}$ & $20~\text{meV}$  & $20~\text{meV}$ \\
$\omega_{2}$ & $20~\text{meV},~ 22.5~\text{meV}$  & $20~\text{meV},~ 22.5~\text{meV}$ \\
$\tilde\lambda_{i}$ & $20~\text{meV}$  & $4~\text{meV}$ \\ 
$\varepsilon_{0}$ & $0$ & $0$\\
$t_{{0}}$ & $-0.2~\text{eV}$ & $-0.2~\text{eV}$\\
$ D_{\mathrm{u}}^{\pd}, D_{\text{c}}^{(1)}, D_{\text{c}}^{(2)}$ & $1~\text{eV}$ & $1~\text{eV}$ \\
$ m_i$ & $5~\mathrm{amu}$ &$5~\mathrm{amu}$ \\
$k_{\mathrm {B}}T$ & $25.8~\text{meV}$ & $25.8~\text{meV}$\\
$\Gamma_{\alpha}$ & $1~\text{eV}$ & $1~\text{eV}$\\
\hline
\end{tabular}
\caption{Two sets of parameters used for the analysis in this work. The only difference between both sets are the electronic-vibrational couplings, $\tilde \lambda_i$.}
\label{tab: paras_table}
\end{table}


Previous studies of purely harmonic models with similar parameters have predicted strong vibrational instabilities in the case of degenerate vibrational modes. For example, in Ref.~\cite{rtr4-xnny}, a large vibrational excitation was observed at relatively low bias voltages of $\Phi=0.3~\mathrm{V}$. Consequently, we are particularly interested in whether a similar effect can also occur in a dissociative model. Specifically, we examine the case where the effective frequency of mode $2$ is degenerate with the frequency of the harmonic potential of mode $1$: $\omega_{2} = \omega_{1}$. If the mechanism of vibrational instability had a significant effect on the dynamics, then it could increase the dissociation probability beyond that of normal Joule heating. 

\begin{figure} 
     \centering
     \includegraphics[width=\columnwidth, trim=0 0 0 0, clip]{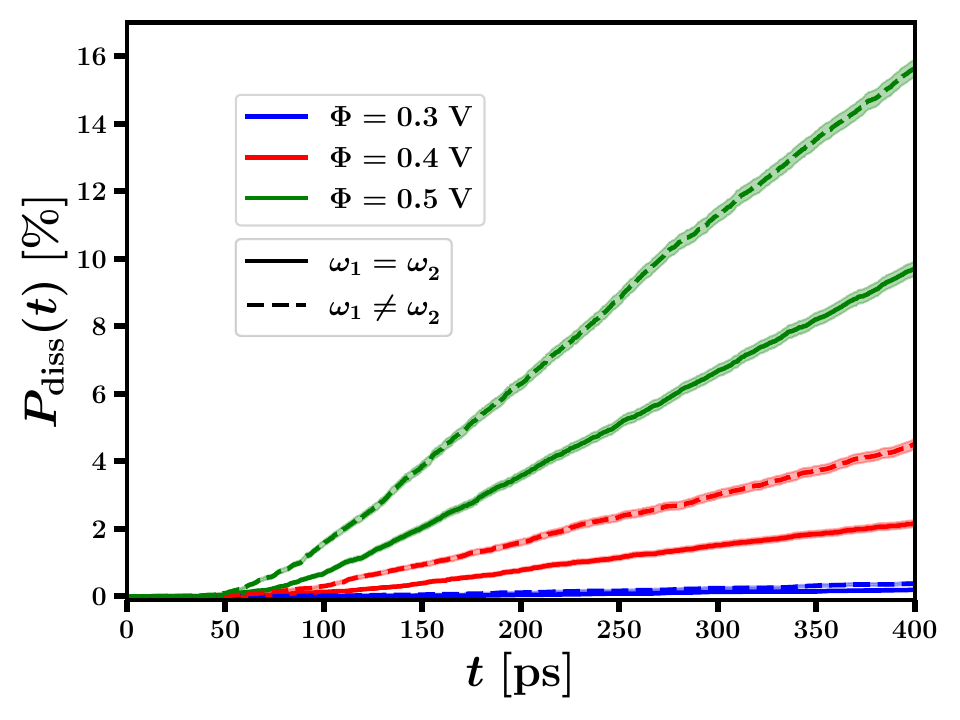}
     \caption{Dissociation probability $P_{\mathrm{diss}}$ over time for different voltages and  the two cases  $\omega_1=\omega_{2}$ and  $\omega_1\neq\omega_{2}$. For the case of $\omega_1\neq\omega_{2}$, we set $\omega_1=20~\mathrm{meV}$ and $\omega_2=22.5~\mathrm{meV}$. Other parameters are listed in Set (1) in Tbl.~\ref{tab: paras_table}. The dissociation threshold is $x^{\mathrm{diss}}_2=10~\mathrm{\mathring{A}}$. The number of initial trajectories for each curve is $N_{\mathrm{traj}}=20000$. The standard deviation is marked by the shaded area around each curve.}
     \label{fig: dissociation_parameter_set_1}
 \end{figure}

To this end, in Fig.~\figref{fig: dissociation_parameter_set_1}, we show the dissociation probability as a function of time for various voltages and the degenerate, $\omega_1=\omega_{2}$, and nondegenerate, $\omega_1\neq\omega_{2}$, cases. Parameters have been taken from Set (1). Fig.~\figref{fig: dissociation_parameter_set_1} demonstrates that the dissociation probability at time $t$ increases with increasing voltage. Counterintuitively, for all voltages considered here, it appears that the dissociation probability in the degenerate case is consistently smaller than in the nondegenerate. We discuss this in further detail below.

Although not shown here, the dissociation probability for Set (2) remains below $1\%$ over the timescale considered in Fig.~\figref{fig: dissociation_parameter_set_1}, even at large bias voltages, $\Phi = 2~\mathrm{V}$. This indicates that the junction is more stable for smaller electronic-vibrational couplings. Furthermore, for Set (2), the dissociation probabilities do not differ significantly between the cases of degenerate and nondegenerate degenerate  modes.

The behavior shown in Fig.~\figref{fig: dissociation_parameter_set_1} is in clear contrast to the mechanism of vibrational instability in multi-mode degenerate harmonic systems \cite{L2010, L2011,L2012,rtr4-xnny}, such as in Eq.\eqref{eq: nuclear_potential_harm} and Eq.\eqref{eq: hamiltonian_2x2_harm}. While in the harmonic case, vibrational dynamics is highly unstable for $\omega_1 = \omega_{2}$, it appears that higher excitation and hence higher dissociation occurs for nondegenerate modes in this anharmonic system. Consequently, it appears that, the original effect leading to a vibrational instability for harmonic systems vanishes when at least one of the nuclear potentials is a Morse potential.

\begin{figure} 
     \centering
     \includegraphics[width=\columnwidth, trim=10 7 10 10, clip]{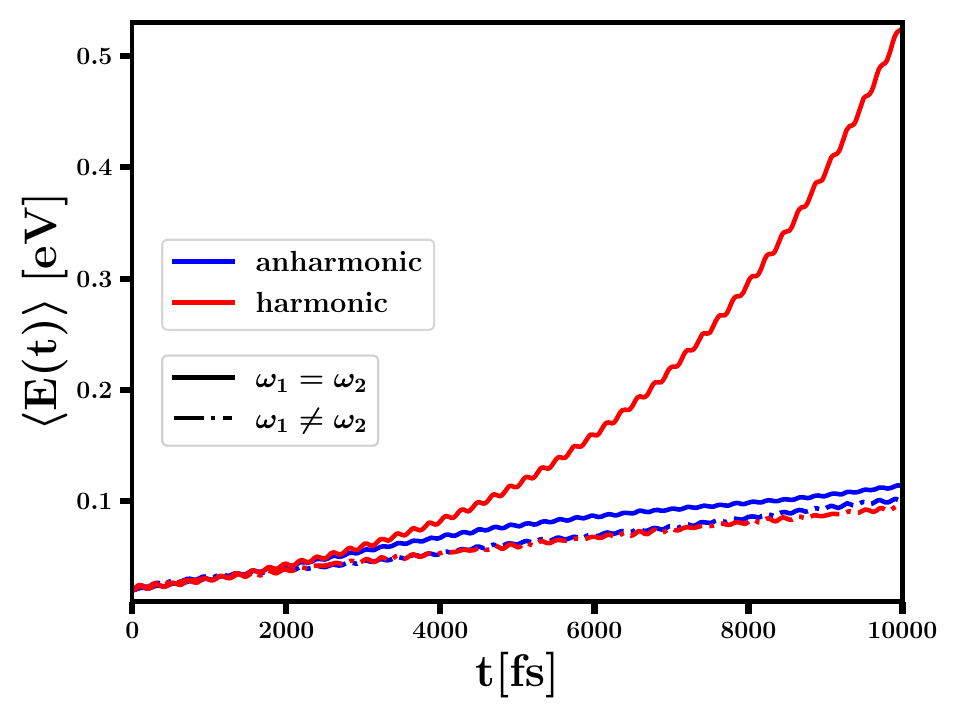}
     \caption{Total vibrational energy $\langle E\rangle =\langle E_1\rangle+\langle E_2\rangle$ over time for the anharmonic and harmonic system and the two respective cases $\omega_1=\omega_{2}$ and  $\omega_1\neq\omega_{2}$.
     For the case of $\omega_1\neq\omega_{2}$, we set $\omega_1=20~\mathrm{meV}$
     and $\omega_2=22.5~\mathrm{meV}$. Other parameters are the same as in parameter set (1) in Tbl.~\ref{tab: paras_table}. The voltage is $\Phi=0.5 ~\mathrm{V}$.} 
     \label{fig: Energy_over_time_harmonic_anharmonic}
 \end{figure}

To investigate the harmonic and anharmonic differences in more detail, in Fig.~\figref{fig: Energy_over_time_harmonic_anharmonic} we compare the total vibrational energy of this anharmonic system to the total vibrational energy obtained in the harmonic limit. In the harmonic limit, the vibrational dynamics differ significantly between the degenerate and nondegenerate cases. In the degenerate case, the energy of both modes increases rapidly over time, eventually leading to a vibrational instability. Conversely, in the nondegenerate case, the vibrational energy increases substantially slower than in the degenerate case, in agreement with Ref.~\cite{rtr4-xnny,L2012,L2011,L2010}. In the anharmonic limit, however, the degeneracy of $\omega_1$ and $\omega_{2}$ only leads to a marginal difference in the vibrational dynamics of both systems. Furthermore, the growth of $\langle E \rangle(t)$ in the degenerate and nondegenerate cases is significantly slower than in the degenerate harmonic system, closely resembling the behavior of the nondegenerate harmonic case. The insensitivity of the vibrational dynamics to mode degeneracy in the anharmonic system implies that the dissociation probability of the junction is also not enhanced by $\omega_1$ and $\omega_{2}$ being degenerate, as shown in Fig.~\figref{fig: dissociation_parameter_set_1}.

Next, we discuss why the vibrational dynamics of the anharmonic model is largely unaffected by mode degeneracy. The harmonic approximation to the anharmonic potentials, $U^{\mathrm{}}_{\mathrm{u},2}(x_2)$ and $U^{(m)}_{\mathrm{c},2}(x_2)$, is obtained via a Taylor expansion around the expansion point $x^{0}_2$. In the vicinity of this point, the harmonic approximation is accurate, $U^{\mathrm{harm}}_{\mathrm{u},2}(x_2) \approx  U^{\mathrm{}}_{\mathrm{u},2}(x_2)$, and the vibrational dynamics essentially experiences two harmonic potentials with frequencies $\omega_1$ and $\omega_{2}$. However, once the molecular trajectory moves away from $x^{\text{0}}_2$, the harmonic approximation becomes less accurate and the effective frequency $\omega_{2}$ does not reflect the true restoring force of the Morse potential. Indeed, one observes in Fig.~\figref{fig: potentials_parameter_set_1} that the effective frequency $\omega_{2} $ of the Morse potential in the direction of dissociation will be smaller than the bottom of the well, effectively detuning the oscillators away from $x^{0}_2$.  

Since the mechanism leading to vibrational instabilities is highly sensitive to frequency detuning between the modes, the effect vanishes once the vibrational dynamics leaves the vicinity of $x^{\text{0}}_2$. Consequently, there is no effect on the dissociation dynamics, because, although the vibrational motion can be excited around the bottom of the well, the dominant contribution once it nears $x_{\text{diss}}$ is still Joule heating, which does not differ significantly between the degenerate and nondegenerate cases. 

\begin{figure} 
     \centering
     \includegraphics[width=\columnwidth, trim=10 7 10 10, clip]{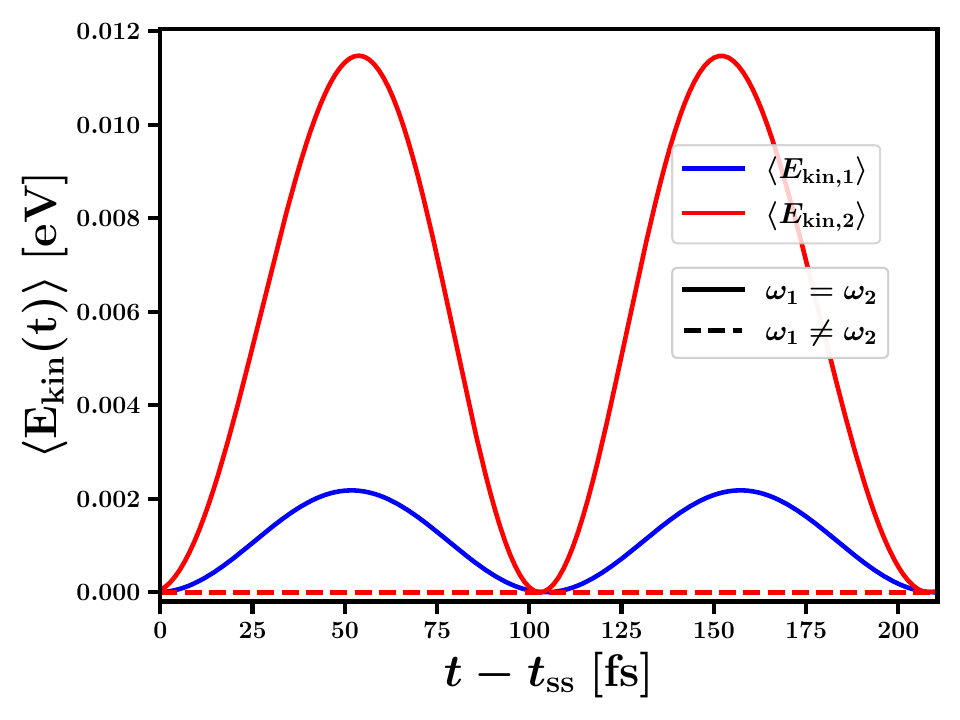}
     \caption{Kinetic energies $\langle E_{\mathrm{kin},1} \rangle$ and $ \langle E_{\mathrm{kin},2} \rangle$ of the individual vibrational modes over time for the anharmonic system.
     At the time $t_{\mathrm{ss}}$, the system has reached a limit cycle and the average kinetic energy stops changing over time.
     Shown are the two cases $\omega_1=\omega_{2}$, and  $\omega_1\neq\omega_{2}$. 
     The trajectory has been initialized with the initial conditions $(x_1,x_2)=(0,0)$ and $(p_1,p_2)=(0,0)$. The parameters are the same as for Fig.~\figref{fig: dissociation_parameter_set_1}, however the stochastic force is removed: $\mathbf{f}(t) = \boldsymbol{0}$. The voltage is $\Phi= 0.5~ \text{V}$. }
     \label{fig: single_trajectory_degen_nondegen}
\end{figure}
\begin{figure} 
     \centering
     \includegraphics[width=\columnwidth, trim=10 7 10 10, clip]{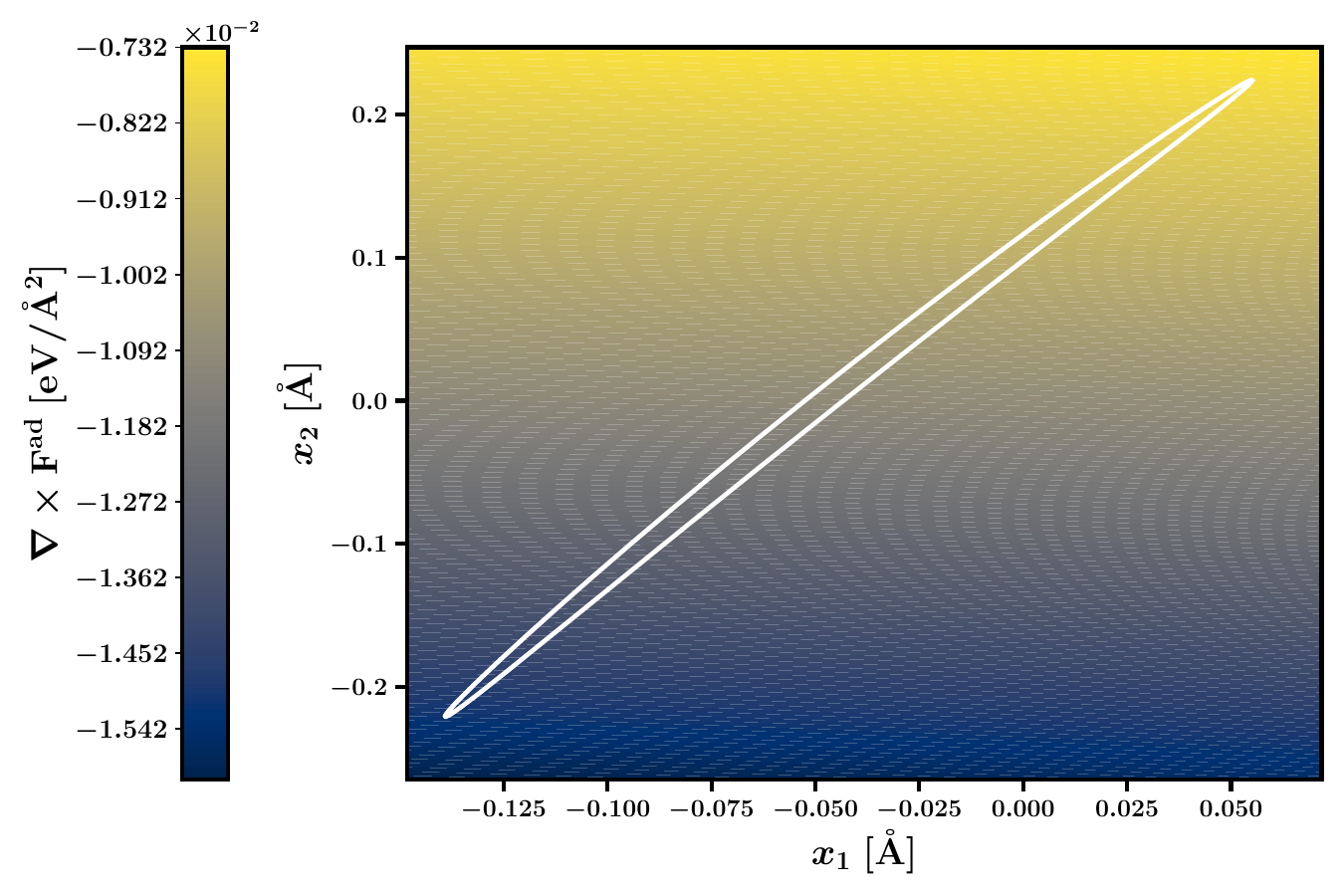}
     \caption{Trajectory in the $(x_1,x_2)$ plane after reaching a limit cycle in white for the anharmonic model and the case $\omega_1=\omega_{2}$. The random force has been removed, $\mathbf{f}(t) = \boldsymbol{0}$. The bias voltage is $\Phi=0.5~\mathrm{V}$. The background shows the curl of the force field.}
     \label{fig: trajectory_on_curl_grid_anharmonic_degen_5lambda_0.5v}
 \end{figure}

This is also observable in Fig.~\figref{fig: single_trajectory_degen_nondegen}, where remnants of the effect are visible. Here, the time evolution of the kinetic energy of a single trajectory for both $\omega_1 = \omega_{2}$ and $\omega_1 \neq \omega_{2}$ for the full anharmonic Morse potential is shown. To remove the perturbative influence of the stochastic force, we have also set the stochastic force to zero: $\mathbf{f}(t)=0$. Since the trajectories do not equilibrate without a stochastic force, we plot the part of the trajectory once it reaches a limit cycle after time $t_{\text{ss}}$. If $\omega_1=\omega_{2}$, the molecular trajectory in the vicinity of $x^{\text{0}}_2$ follows an ellipse with fixed rotational direction in the $(x_{1},x_{2})$ plane, as shown in Fig.~\figref{fig: trajectory_on_curl_grid_anharmonic_degen_5lambda_0.5v}. Here, one also observes that the adiabatic electronic force field is nonconservative: $\nabla \times \bm{F}(\bm{x}) < 0$. 

Consequently, in this region, the trajectory gains energy from $\bm{F}(\bm{x})$ that can compensate energy loss due to the electronic friction force. Although it is not shown here, the electronic friction tensor is positive definite for the entire vibrational coordinate range, such that it has a purely dissipative effect. Therefore, for degenerate vibrational modes, the vibrational kinetic energy does not decay to zero, as can be seen in Fig.~\figref{fig: single_trajectory_degen_nondegen}. Despite these persistent oscillations, the overall vibrational excitation without the stochastic force is still small, as the effect vanishes when the trajectory moves too far away from $ x^\mathrm{0}_2$, such that electronic friction dominates and effectively capping the maximum kinetic and potential energy. 

In contrast, in the nondegenerate case, the molecular trajectory does not follow a fixed rotational direction and will, in total, not gain energy from the force field. For clarity, this has not been shown in Fig.~\figref{fig: trajectory_on_curl_grid_anharmonic_degen_5lambda_0.5v}, as it obscures the elliptical trajectory for $\omega_{1} = \omega_{2}$. Furthermore, because the random force has been artificially turned off, there is no compensation for dissipation due to the electronic frictional force and the kinetic energy decays to zero over time, as shown in Fig.~\figref{fig: single_trajectory_degen_nondegen}. 

\subsection{Influence of Quartic Anharmonicity on the Vibrational Dynamics}\label{subsec: quartic_vib_energy}

In the previous section, it was shown that the mechanism of vibrational instability present for degenerate vibrational frequencies and harmonic vibrational modes largely disappears when one mode is replaced by a Morse potential, and has no effect on the dissociative dynamics. However, as shown in Fig.~\figref{fig: single_trajectory_degen_nondegen}, remnants of the mechanism are still present near the bottom of the potential wells. Given that realistic molecular potentials are generally anharmonic, this raises several questions. First, will introducing an anharmonicity always destroy the mechanism of vibrational instability? Second, given that a remnant of the effect remains, are there parameters that can magnify the vibrational instability in anharmonic models? 

To explore these questions, in this section, we investigate the degree of anharmonicity a system is allowed to exhibit before the signature of the vibrational instability vanishes. In order to simplify the effect of the anharmonicity, we will not use a Morse potential, instead taking the harmonic system introduced in Eq.\eqref{eq: hamiltonian_2x2_harm} and Eq.\eqref{eq: nuclear_potential_harm} and adding a quartic contribution to $U^{\mathrm{}}_{\mathrm{u},2}(x_2)$,
\begin{equation}
 U^{\mathrm{}}_{\mathrm{u},2}(x_2) =   \frac{1}{2} m_2 \omega_2^2x_2^2 +\alpha x_2^4.
\end{equation}
This allows us to rigorously control the strength of the anharmonicity via a single parameter, $\alpha$. 

\begin{figure*} 
\begin{center}
     \centering     
     \includegraphics[width=1.0\textwidth, trim=0 0 0 0, clip]{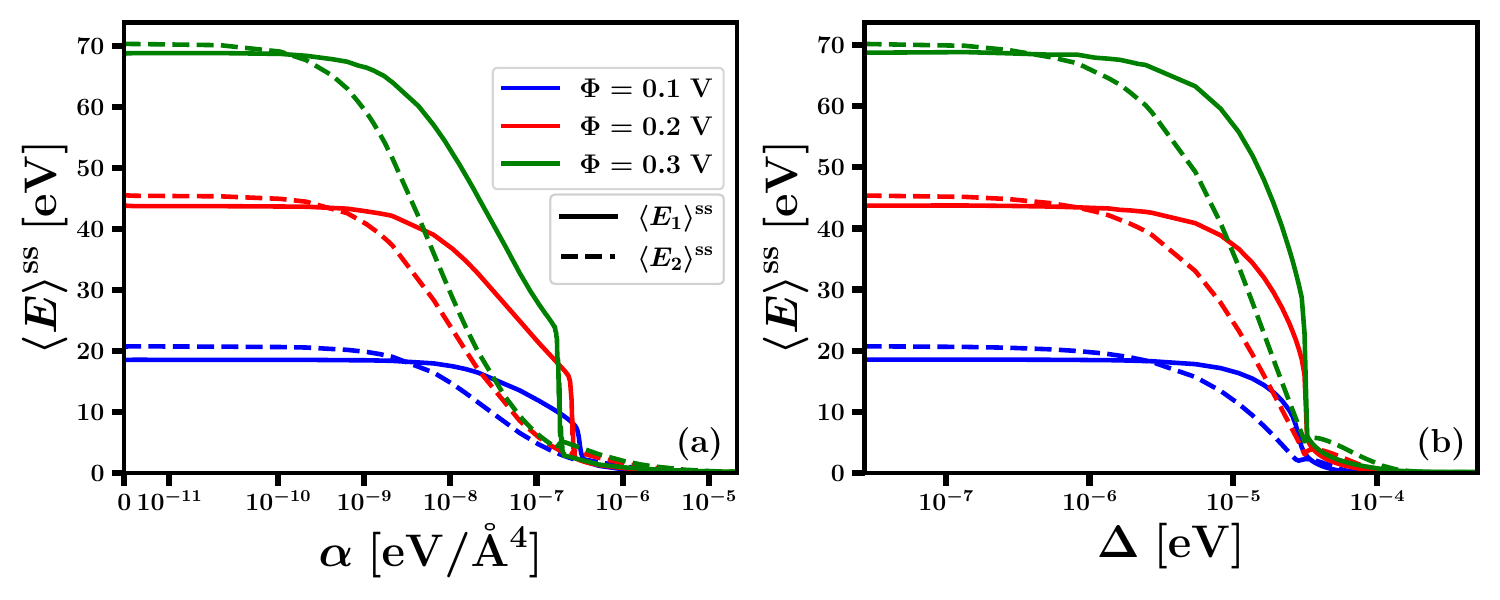}
     \phantomsubfloat{\label{fig: vib_energy_quartic_5lambda}}
     \phantomsubfloat{\label{fig: vib_energy_harmonic_5lambda}}
     \vspace{-.80cm}
     \caption{(a) Steady state vibrational energy $\langle E_1 \rangle^{\mathrm{ss}}$ and $\langle E_2 \rangle^{\mathrm{ss}}$ of the individual modes in the quartic system for different values of the anharmonicity parameter $\alpha$ and different bias voltages. The frequencies are $\omega_1=\omega_2=20~\mathrm{meV}$, all other parameters are the same as in parameter set (1) listed in Tbl.~\ref{tab: paras_table}. (b) Steady state vibrational energy of the individual modes for the harmonic system for different values of frequency difference $\Delta$. The frequencies are $\omega_1 = 20~\mathrm{meV}$ and $\omega_2 = \omega_1+ \Delta$. For all frequencies, the electronic vibrational couplings are chosen such that $\lambda_{1/2} = \tilde \lambda_{1} \sqrt{m_{1}\omega_1}$. 
     All other parameters are the same as for the quartic system.}
     \label{fig: vib_energy_quartic_harmonic_5lambda}
     \end{center}
 \end{figure*}   

In Fig.~\figref{fig: vib_energy_quartic_5lambda}, we show the vibrational energy for different values of the anharmonicity parameter $\alpha$ and different bias voltages for the quartic system introduced above. Note that we show the steady-state vibrational energy, as this can be reached for the bound quartic potentials. Based on Fig.~\figref{fig: vib_energy_quartic_5lambda}, we can distinguish three regimes in the vibrational dynamics for all voltages considered. In the first regime, where $0<\alpha<10^{-9}~\frac{\mathrm{eV}}{\mathring{\mathrm{A}}^4}$, we observe a clear vibrational instability. Here, the vibrational energy in the steady state reaches unrealistically high values already at very low bias voltages. Note that a realistic junction would already break before reaching such high vibrational energies. The fact that we are able to observe such high energies is due to the non-dissociative potentials of our model.   

The second regime, where $10^{-9}~\frac{\mathrm{eV}}{\mathring{\mathrm{A}}^4}<\alpha<10^{-6}~\frac{\mathrm{eV}}{\mathring{\mathrm{A}}^4}$, marks the suppression of this instability. While the vibrational energy in this regime is still substantial, the energy distribution among both vibrational modes changes significantly and the energy difference between both modes becomes larger. The energy of vibrational mode $\langle E_1 \rangle^{\mathrm{ss}}$ only starts decreasing for slightly larger $\alpha$ than  $\langle E_2 \rangle^{\mathrm{ss}}$, as the latter contains the quartic contribution. 

We note that in this regime, the system does not exhibit a unique steady state for certain values of $\alpha$. For the latter mentioned points, the stochastic force pushes individual trajectories to one of two steady states, which differ significantly in their vibrational energy. This is also connected to the steep decrease in vibrational energy between $10^{-7}~\frac{\mathrm{eV}}{\mathring{\mathrm{A}}^4}<\alpha<10^{-6}~\frac{\mathrm{eV}}{\mathring{\mathrm{A}}^4}$. A detailed analysis of the vibrational dynamics within this regime lies beyond the scope of the main text and does not impact the following analysis. We therefore defer it to the Appendix~\ref{sec: Appendix}. 

The third regime, where $\alpha>10^{-6}~\frac{\mathrm{eV}}{\mathring{\mathrm{A}}^4}$ marks
the stable regime where the signature of the instability vanishes. Here, the vibrational energy is significantly smaller than in the other two regimes and the junction is essentially stable.

The results shown here indicate that adding even a small anharmonicity to one of the harmonic potentials destroys the mechanism leading to a vibrational instability observed in degenerate harmonic models. This observation is similar to our findings in Ref.~\cite{rtr4-xnny}, in which it was shown that the instability also vanishes in harmonic systems upon slightly detuning the vibrational frequencies: $\Delta = \omega_2-\omega_1 \neq 0$. To connect to these ideas, in Fig.~\figref{fig: vib_energy_harmonic_5lambda}, we also show the vibrational kinetic energy of the harmonic approximation to the quartic model as a function of the manual detuning parameter $\Delta$. 

Here, we observe a similar dependence of $\langle E_{i} \rangle^{\text{ss}}$ on $\Delta$ as it had to the anharmonicity parameter $\alpha$ in the quartic system. Specifically, for the harmonic system with manual detuning, we can also distinguish three different regimes of the vibrational dynamics. The regime of instability $(\Delta< 10^{-6}~\mathrm{eV})$, the transition regime in which the instability is suppressed $(10^{-6}~\mathrm{eV}<\Delta< 10^{-4}~\mathrm{eV})$ and the regime where the dynamics become stable $(\Delta> 10^{-4}~\mathrm{eV})$. 

As we observe in Fig.~\figref{fig: vib_energy_quartic_harmonic_5lambda}, increasing the anharmonicity parameter $\alpha$ for the quartic system and increasing the frequency difference in the harmonic system have a similar impact on the vibrational dynamics of the respective system. However, while the results show a qualitative similarity, it seems like the signature of the instability is even more sensitive to anharmonicity than to the frequency difference in the harmonic system. In the following, we explore this behavior in further detail.

\begin{figure} 
     \centering
     \includegraphics[width=\columnwidth, trim=10 7 10 10, clip]{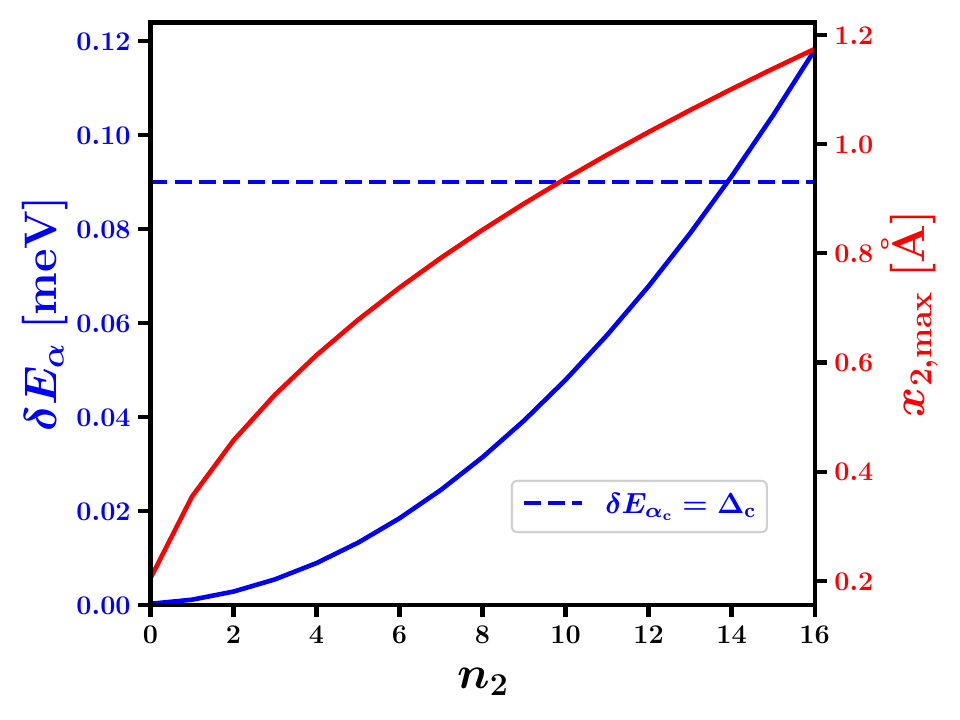}
     \caption{Energy difference $\delta E_{\alpha}$ between the vibrational modes caused by the quartic term, and the classical turning point $x_{2,\mathrm{max}}$ as a function of the occupation number $n_2$.  }
     \label{fig: comparison_enegery_difference_quartic_harmonic}
 \end{figure}

First, we observe the values of $\alpha_{\mathrm{c}}$ and $\Delta_{\mathrm{c}}$ that mark the onset of the instability in Fig.~\figref{fig: vib_energy_quartic_5lambda} and Fig.~\figref{fig: vib_energy_harmonic_5lambda}, respectively. The values of $\alpha_\mathrm{c}$ and $\Delta_\mathrm{c}$ are determined by considering the smallest values of $\alpha$ and $\Delta$ where the vibrational energy in one of the modes exceeds $E_\mathrm{c}=2~\mathrm{eV}$ at a voltage of $\Phi=0.3~\mathrm{V}$, which corresponds to values of $\alpha_{\mathrm{c}}=1\cdot 10^{-6}~\mathrm{\frac{eV}{\mathring{A}^4}}$ and $\Delta_{\mathrm{c}}=9\cdot 10^{-5}~\mathrm{eV}$. Once the vibrational energy in one of the modes exceeds $E_\mathrm{c}$, we assume the junction to exhibit a vibrational instability. Note that, although the exact choice of the vibrational energy marking the onset of an instability is somewhat arbitrary, different choices would not qualitatively impact the following analysis.  

Next, we calculate the energy detuning between the vibrational modes induced by the anharmonicity via first-order perturbation theory, in which $\alpha$ is treated as a small parameter:
\begin{equation}
    \delta E_{\alpha} = \frac{3\alpha}{4m^2_2\omega^2_2}\left(2n_2+2n_2^2 +1\right),
\end{equation}
where $n_2$ is the occupation number of the vibrational mode corresponding to coordinate $x_2$. Note that, although we treat the vibrational modes classically, one can still define a classical harmonic analogue to the occupation number via: $n_{i} = \langle E_{i} \rangle / \omega_{i}$. 

For the  harmonic system, the energy difference between both oscillators is simply
\begin{equation}
    \delta E_{\Delta} = \Delta. 
\end{equation}
By setting $\delta E_{\alpha_\mathrm{c}}=\delta E_{\Delta_\mathrm{c}}$ and using the values of $\alpha_{\mathrm{c}} $ and $\Delta_{\mathrm{c}}$ marking the onset of instability, it is possible to determine the occupation number for which the anharmonicity and the frequency detuning lead to the same energy difference between both modes.

Moreover, by considering the classical turning point of the harmonic oscillator,
\begin{equation}
    x_{2,\mathrm{max}} = \pm \sqrt{\frac{2 \left( n_2+\frac{1}{2} \right)}{m_2\omega_2}},
\end{equation}
we can determine the position at which the energy difference caused by the quartic term becomes larger than the energy difference caused by the frequency detuning. In Fig.~\figref{fig: comparison_enegery_difference_quartic_harmonic}, we show $\delta E_{\alpha_\mathrm{c}}$ and $x_{2,\mathrm{max}}$ for different values of $n_2$. The point at which $\delta E_{\alpha_\mathrm{c}}=\delta E_{\Delta_\mathrm{c}}$ is at a relatively low vibrational occupation, $ n_2 \approx 14$, which corresponds to $x_{2,\mathrm{max}}\approx \pm 1.1~\mathrm{\mathring{A}}$. 

Consequently, we observe that even for relatively small deviations from the bottom of the potential well, $|x_2| > 1.1~\mathrm{\mathring{A}}$, the quartic anharmonicity induces a larger energy deviation of both modes than the manual detuning $\Delta$ that has already been shown to destroy the mechanism of vibrational instability \cite{rtr4-xnny}. As a result, for $\alpha>\alpha_{\mathrm{c}}$, the mechanism leading to the instability is suppressed for such small-amplitude motion that it has no influence on the steady-state vibrational excitation. 

The results presented in this section show that even a small anharmonicity in one of the nuclear potentials destroys the effect leading to a vibrational instability in degenerate harmonic systems. Given that the anharmonicity in this section was a simple quartic term with relatively small anharmonicity parameter, it is unclear whether this mechanism of vibrational instability would be observable in more realistic molecular nanojunction systems, as even molecules containing degenerate vibrational modes are rarely purely harmonic. Indeed, in Sec.~\ref{subsec: diss_dynamics}, where we included more realistic Morse potentials, the dissociation dynamics showed no evidence of the instability. 

\subsection{Influence of Anharmonicity on Steady State Electric Current}\label{subsec: quartic_current}

So far, the discussion was centered on the impact of anharmonicity on the vibrational dynamics. We now turn to its effect on the steady-state electric current. We consider the same model with the same parameters as in Sec.~\ref{subsec: quartic_vib_energy}. 

\begin{figure} 
     \centering
     \includegraphics[width=\columnwidth, trim=0 0 0 0, clip]{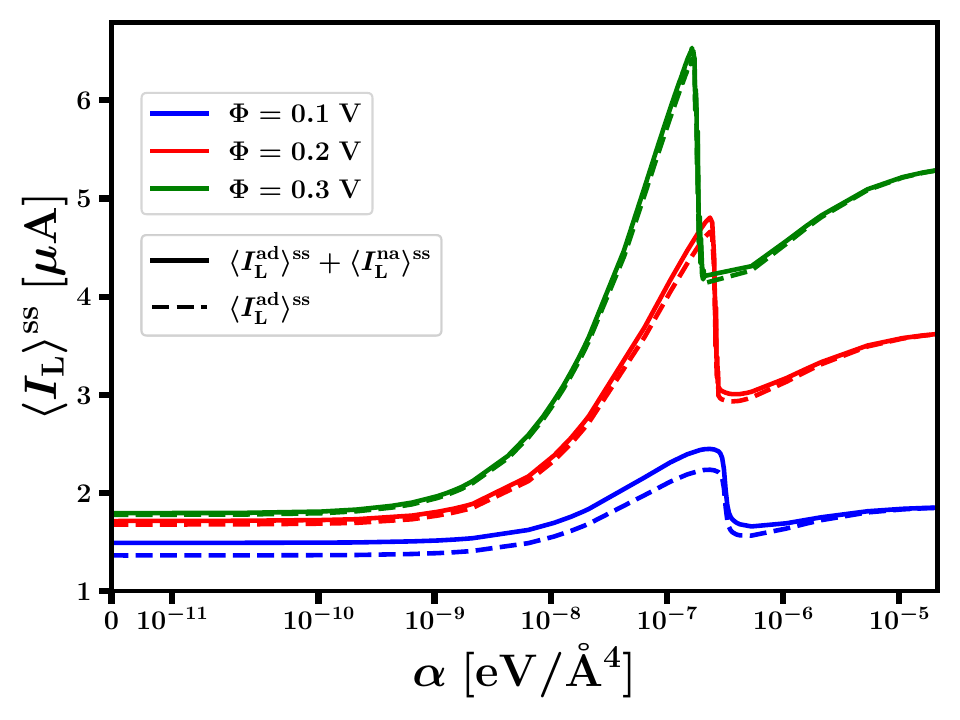}
     \caption{Adiabatic current and its first-order nonadiabatic correction $\langle I_{\mathrm{L}}^{\mathrm{ad}} \rangle^{\mathrm{ss}} +\langle I_{\mathrm{L}}^{\mathrm{na}} \rangle^{\mathrm{ss}}$ and the purely adiabatic current $ \langle I_{\mathrm{L}}^{\mathrm{ad}} \rangle^{\mathrm{ss}}$ for different values of the anharmonicity parameter $\alpha$ and different bias voltages. }
     \label{fig: current_quartic}
 \end{figure}

Fig.~\figref{fig: current_quartic} shows the adiabatic current through the left lead, $ \langle I_{\mathrm{L}}^{\mathrm{ad}} \rangle^{\mathrm{ss}}$, and its first-order nonadiabatic correction $\langle I_{\mathrm{L}}^{\mathrm{ad}} \rangle^{\mathrm{ss}} +\langle I_{\mathrm{L}}^{\mathrm{na}} \rangle^{\mathrm{ss}}$ for the same range of the anharmonicity parameter $\alpha$ as in Fig.~\figref{fig: vib_energy_quartic_5lambda} and different bias voltages. For all voltages shown here, the adiabatic contribution to the current shows a strong signature for $10^{-9}~\frac{\mathrm{eV}}{\mathring{\mathrm{A}}^4}<\alpha<10^{-6}~\frac{\mathrm{eV}}{\mathring{\mathrm{A}}^4}$, which is the same range of $\alpha$ for which the vibrational energy drops significantly in Fig.~\figref{fig: vib_energy_quartic_5lambda}. For increasing $\alpha$, $\langle I_{\mathrm{L}}^{\mathrm{ad}} \rangle_{\mathrm{}}^{\mathrm{ss}} $ first increases reaching its maxima around  $2 \cdot10^{-7}~\frac{\mathrm{eV}}{\mathring{\mathrm{A}}^4}$, before it decreases over a small ranger of $\alpha$ and starts increasing again. 

\begin{figure} 
     \centering
     \includegraphics[width=\columnwidth, trim=10 7 10 10, clip]{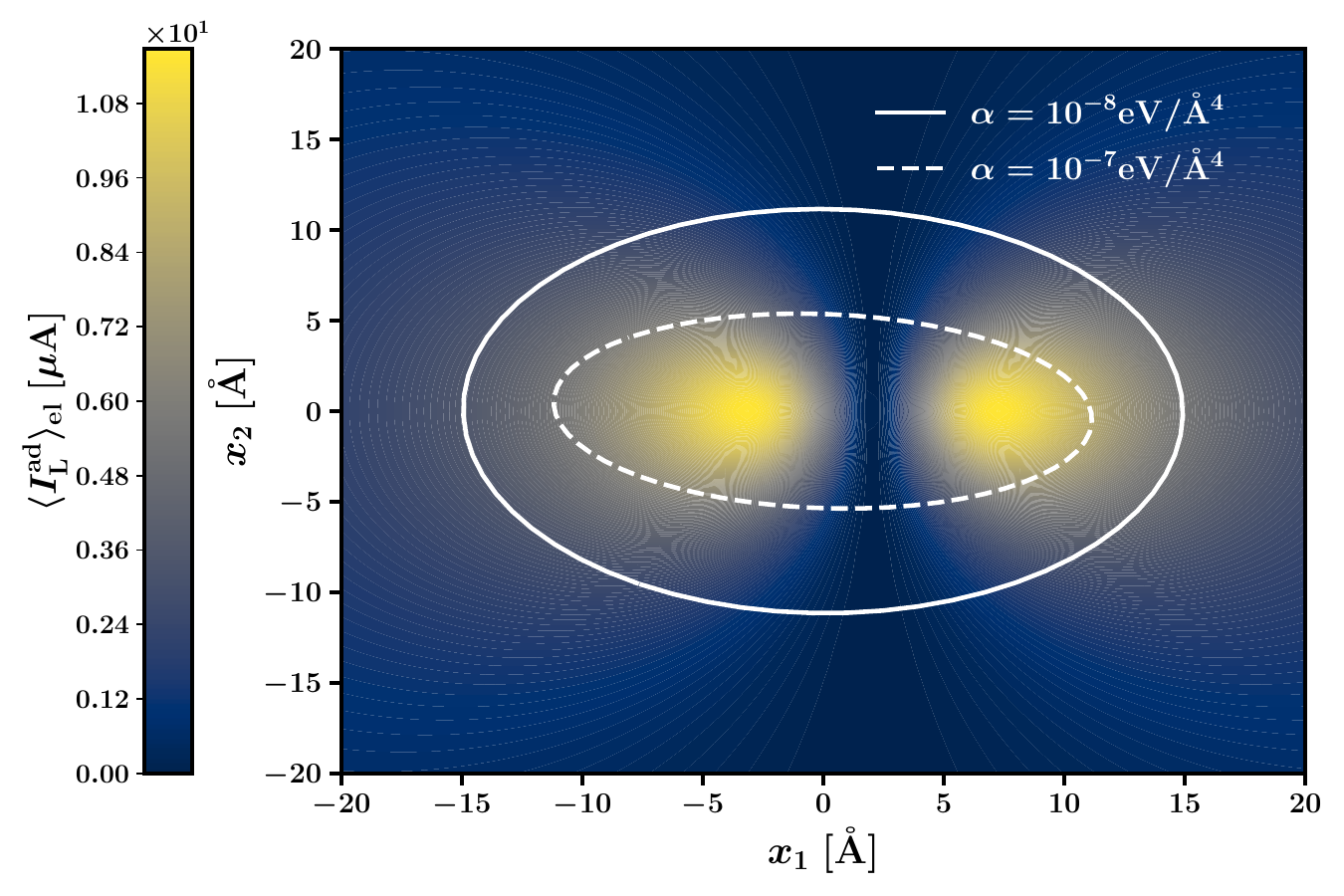}
     \caption{Trajectories in the $(x_1,x_2)$ plane after reaching a limit cycle for different values of $\alpha$. The random force has been removed, $\mathbf{f}(t) = \boldsymbol{0}$. The bias voltage is $\Phi=0.3~\mathrm{V}$. The background shows the adiabatic current, $\langle I_{\mathrm{L}}^{\mathrm{ad}}\rangle_{\mathrm{el}} (\bm{x}) $, as given in Eq.\eqref{eq: instantaneous adiabatic current}.}
     \label{fig: trajectories_on_current_grid_5lambda_0.3v}
 \end{figure}

This signature in the adiabatic current can be attributed to the fact that, between $10^{-9}~\frac{\mathrm{eV}}{\mathring{\mathrm{A}}^4}<\alpha<10^{-6}~\frac{\mathrm{eV}}{\mathring{\mathrm{A}}^4}$, the vibrational dynamics change significantly, redistributing the vibrational energy among both modes, as we showed in Fig.~\figref{fig: vib_energy_quartic_5lambda}. Therefore, within this range of $\alpha$, the molecular trajectory also changes considerably. To illustrate this, we show the molecular trajectory in the $(x_1,x_2)$ plane in Fig.~\figref{fig: trajectories_on_current_grid_5lambda_0.3v} exemplarily for two values of $\alpha$. 
For $\alpha=10^{-8}~\frac{\mathrm{eV}}{\mathring{\mathrm{A}}^4}$, the radius of curvature is larger than for $\alpha=10^{-7}~\frac{\mathrm{eV}}{\mathring{\mathrm{A}}^4}$, since the system possesses more vibrational energy, as shown in Fig.~\figref{fig: vib_energy_quartic_5lambda}. Moreover, considering the magntitude of the adiabatic current,  $\langle I_{\mathrm{L}}^{\mathrm{ad}}\rangle_{\mathrm{el}} (\bm{x}) $, we see that for $\alpha=10^{-7}~\frac{\mathrm{eV}}{\mathring{\mathrm{A}}^4}$, on average,  the trajectory intersects the areas of the coordinate grid where $\langle I_{\mathrm{L}}^{\mathrm{ad}}\rangle_{\mathrm{el}} (\bm{x}) $ has a larger magnitude more often than for $\alpha=10^{-8}~\frac{\mathrm{eV}}{\mathring{\mathrm{A}}^4}$. 
This change in the radius of curvature and the shape of the trajectory in the steady-state leads to the observed non-monotonous behavior in the current for different values of $\alpha$.

Adding the first-order nonadiabatic correction to the current increases the overall current slightly for all voltages considered. This increase in magnitude can be explained considering the formula for the nonadiabatic contribution to the current in Eq.\eqref{eq: instantaneous nonadiabatic current}. As in regimes of the vibrational instability the nuclear momenta are substantially large, and the 
nonadiabatic correction is proportional to the individual momenta, the nonadiabatic correction to the current becomes relevant.

For increasing bias voltage on the other hand, the nonadiabatic correction becomes less relevant.
This can be attributed to the fact that the radius of curvature of the molecular trajectory increases for increasing bias, since the trajectory gains more energy from the force field (see Fig.~\figref{fig: vib_energy_quartic_5lambda}). Once the radius becomes sufficiently large, the trajectory no longer intersects the areas of the coordinate grid where the $I_{\mathrm{L},j}^{\mathrm{na}} (\bm{x})$ have a significant magnitude, similar to what we showed for the adiabatic current in Fig.~\figref{fig: trajectories_on_current_grid_5lambda_0.3v}. 
Thus, for increasing voltage, the nonadiabatic correction to the current decreases. 
Moreover, since for larger $\alpha$, the instability is suppressed, and the nuclear momenta become substantially smaller, the nonadiabatic current becomes less relevant in this regime.

\begin{figure*} 
\begin{center}
     \centering     
     \includegraphics[width=1.0\textwidth, trim=0 0 0 0, clip]{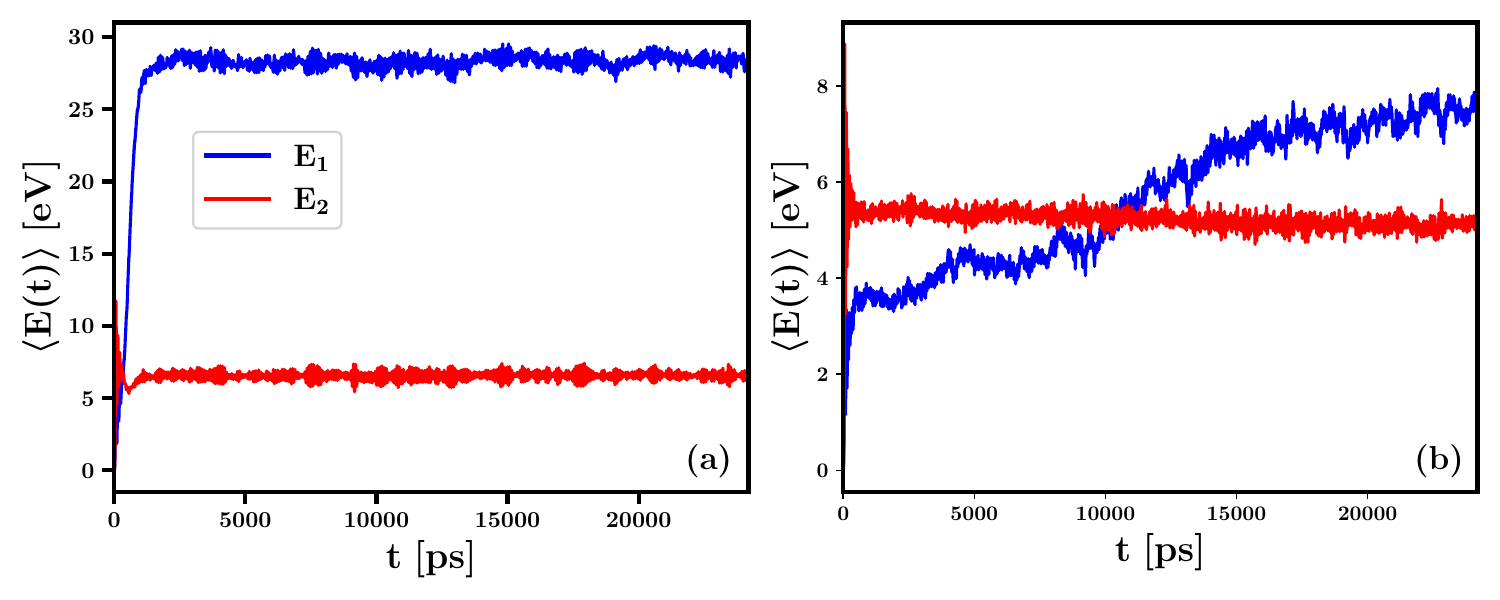}
     \phantomsubfloat{\label{fig: mean_vib_energy_over_time_alpha_1}}
     \phantomsubfloat{\label{fig: mean_vib_energy_over_time_alpha_2}}
     \vspace{-.80cm}
     \caption{(a) Average vibrational energy of the two vibrational modes over time for the system from Sec.~\ref{subsec: quartic_vib_energy} and a value of $\alpha=1\cdot10^{-7} \mathrm{\frac{eV}{\mathring{A}^4}}$. (b) Average vibrational energy over time for $\alpha=1.8\cdot10^{-7} \mathrm{\frac{eV}{\mathring{A}^4}}$. The initial conditions for all trajectories  are  $(x_1,x_2)=(0,0)$ and $(p_1,p_2)=(0,0)$, in contrast to the main text. }
     \label{fig: mean_vib_energy_over_time}
     \end{center}
 \end{figure*}   

\section{Conclusion} \label{sec: Conclusion}

In this work, we investigated the influence of anharmonic nuclear potentials on the vibrational dynamics and the corresponding mechanical stability of multi-mode molecular nanojunctions. This extends previous investigations of vibrational instabilities originating from nonconservative electronic forces, which have so far been restricted to purely harmonic nuclear potentials. 
To simulate the vibrational dynamics in our model systems, we used a mixed quantum-classical approach based on electronic friction and Langevin dynamics. 

First, we investigated the vibrational dynamics in a dissociative model containing one harmonic and one Morse oscillator. We showed that the dissociation dynamics of this model was almost completely unaffected by the degeneracy of the effective frequencies, although remnants of the effect were visible for small regions around the bottom of the potential well. Based on our simulations, we would not expect this mechanism to show a significant effect when measuring dissociation rates in an experiment. 


Next, we investigated whether the specific form and strength of the anharmonicity affects the mechanism of vibrational instability. To this end, we additionally considered a model containing one harmonic and one quartic oscillator. We found that adding even a small quartic term to a harmonic potential destroys this mechanism. Furthermore, we connected these findings to previous work on purely harmonic oscillators, where it was found that the instability vanishes even for small detunings of the two oscillators. Our analysis showed that the effective detuning induced by even very small anharmonicities is much larger than what is needed to break the mechanism and stabilize the junction. Finally, we investigated the influence of the anharmonicity on the steady state current, showing that the steady-state current exhibits a strong signature near the onset of the instability. 

While the analysis in this paper has been restricted to generic models of molecular nanojunctions with anharmonic nuclear potentials, our results indicate that anharmonicity destroys the original effect that leads to vibrational instabilities in harmonic models. Based on our results, it is therefore questionable if instabilities specifically originating from current-induced nonconservative forces could be observed in a molecular junction. Since our results are obtained using a mixed quantum-classical method, benchmark calculations of the full quantum dynamics, for example using the HEOM approach \cite{Tanimura1989, Schinabeck2020,Jin2008, Hrtle2013,Hrtle2018,Schinabeck2016}, could provide a validation of our simulations.

\section*{Acknowledgment}

This work was supported by the Deutsche Forschungsgemeinschaft (DFG) through Research Unit FOR5099, as well as support from the state of Baden-Württemberg through bwHPC and the DFG through Grant No. INST 40/575-1 FUGG (JUSTUS 2 cluster). Moreover, the authors thank Mads Brandbyge, Jing-Tao Lü, Joseph Subotnik, and Riley Preston for helpful discussions.

\appendix

\section{More Detailed Analysis of the Vibrational Dynamics in the Quartic Model} \label{sec: Appendix}

\begin{figure} 
     \centering
     \includegraphics[width=\columnwidth, trim=0 0 0 0, clip]{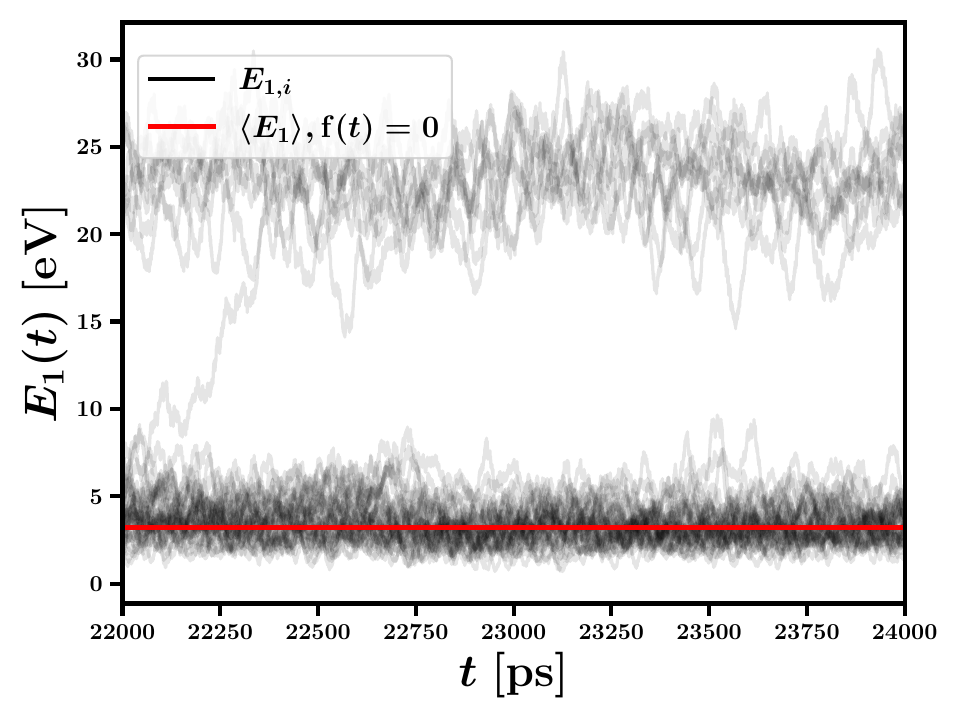}
     \caption{Vibrational energy of individual trajectories over time for mode (1) and $\alpha=1.8\cdot10^{-7} \mathrm{\frac{eV}{\mathring{A}^4}}$. The red curve shows the average vibrational energy for mode (1) within the time considered here obtained by setting $\mathbf{f}(t) = 0$. All trajectories have the same initial condition, $(x_1,x_2)=(0,0)$ and $(p_1,p_2)=(0,0)$.  }
     \label{fig: energy_individual_trajectories_over_time_alpha_2}
 \end{figure}

In this Appendix, we present a more detailed analysis of the vibrational dynamics shown in Fig.~\figref{fig: vib_energy_quartic_5lambda} in Sec.~\ref{subsec: quartic_vib_energy}. In particular, we discuss the occurrence of the steep decrease of the vibrational energy in Fig.~\figref{fig: vib_energy_quartic_5lambda}. We start by showing the expectation values of the vibrational energy over time for two different values of $\alpha$ in Fig~\figref{fig: mean_vib_energy_over_time}. The first value of $\alpha=1\cdot10^{-7} \mathrm{\frac{eV}{\mathring{A}^4}}$ lays distant from the kink in Fig.~\figref{fig: vib_energy_quartic_5lambda}, while the second value of $\alpha=1.8\cdot10^{-7} \mathrm{\frac{eV}{\mathring{A}^4}}$ lays within the region of the kink.

While Fig.~\figref{fig: mean_vib_energy_over_time_alpha_1} shows that within the timescale considered here both modes have reached the vibrational steady state, Fig.~\figref{fig: mean_vib_energy_over_time_alpha_2} shows that the average vibrational energy is still changing over time. While Fig.~\figref{fig: mean_vib_energy_over_time_alpha_2} suggest that the vibrational steady state is not yet reached, we demonstrate in the following that the system exhibits no unique steady state.

In Fig.~\figref{fig: energy_individual_trajectories_over_time_alpha_2} we show the vibrational energy of individual trajectories for mode (1) for $\alpha=1.8\cdot10^{-7} \mathrm{\frac{eV}{\mathring{A}^4}}$. To remove any dependence of the steady state on the initial condition, all trajectories have been initialized with the same initial condition, $(x_1,x_2)=(0,0)$ and $(p_1,p_2)=(0,0)$, in contrast to the main text. The energy of the individual trajectories oscillates around two different mean values, indicating that the system exhibits two steady states. Moreover, when setting $\mathbf{f}(t) = 0$, only one steady state is found with similar energy as in one of the beforehand mentioned steady states when simulating the full vibrational dynamics. We therefore conclude that in the vicinity of the kink in Fig.~\figref{fig: vib_energy_quartic_5lambda}, the stochastic force drives individual trajectories to one of two different steady states. 

Since we do not observe a unique steady state for certain values of $\alpha$, we have to adjust the procedure outlined in Sec.~\ref{subsec: Langevin simulations and Observables} to obtain the steady state observables. 
In particular, we decide to calculate our observables by averaging over both steady states, weighted by the number of trajectories in each of them. 
Therefore, depending on the value of $\alpha$, the number of trajectories in each steady state may change abruptly, which leads to the observed kink in Fig.~\figref{fig: vib_energy_quartic_5lambda}. Note that, as already mentioned in the main text, this concerns only values of $\alpha$ near the kink in Fig.~\figref{fig: vib_energy_quartic_5lambda}, and does not impact the further analysis in the main text. 

\clearpage

\newpage

\bibliography{references.bib}

@article{PhysRevB.107.085419,
  title = {Current-driven collective dynamics of non-Hermitian edge vibrations in armchair graphene nanoribbons},
  author = {Mao, Wen-Hao and Shang, Man-Yu and L\"u, Jing-Tao},
  journal = {Phys. Rev. B},
  volume = {107},
  issue = {8},
  pages = {085419},
  numpages = {9},
  year = {2023},
  month = {Feb},
  publisher = {American Physical Society},
  doi = {10.1103/PhysRevB.107.085419},
  url = {https://link.aps.org/doi/10.1103/PhysRevB.107.085419}
}

@article{Schulze2008,
  title = {Resonant Electron Heating and Molecular Phonon Cooling in Single C60 Junction},
  volume = {100},
  ISSN = {1079-7114},
  url = {http://dx.doi.org/10.1103/PhysRevLett.100.136801},
  DOI = {10.1103/physrevlett.100.136801},
  number = {13},
  journal = {Physical Review Letters},
  publisher = {American Physical Society (APS)},
  author = {Schulze,  G. and Franke,  K. J. and Gagliardi,  A. and Romano,  G. and Lin,  C. S. and Rosa,  A. L. and Niehaus,  T. A. and Frauenheim,  Th. and Di Carlo,  A. and Pecchia,  A. and Pascual,  J. I.},
  year = {2008},
  month = apr 
}

@article{rtr4-xnny,
  title = {Vibrational instabilities in charge transport through molecular nanojunctions: The role of nonconservative current-induced electronic forces},
  author = {M\"ack, Martin and Preston, Riley J. and Thoss, Michael and Rudge, Samuel L.},
  journal = {Phys. Rev. B},
  volume = {112},
  issue = {7},
  pages = {075430},
  numpages = {15},
  year = {2025},
  month = {Aug},
  publisher = {American Physical Society},
  doi = {10.1103/rtr4-xnny},
  url = {https://link.aps.org/doi/10.1103/rtr4-xnny}
}

@article{Lue2019,
journal = {Prog Surf Sci.},
title={Semi-classical generalized Langevin equation for equilibrium and nonequilibrium molecular dynamics simulation},
volume = {94},
number = {1},
pages = {21-40},
year = {2019},
issn = {0079-6816},
doi = {https://doi.org/10.1016/j.progsurf.2018.07.002},
url = {https://www.sciencedirect.com/science/article/pii/S0079681618300200},
author = {Jing-Tao Lü and Bing-Zhong Hu and Per Hedegård and Mads Brandbyge}
}

@article{Christensen2016,
author={Rasmus Bjerregaard Christensen and Jing-Tao Lü and Per Hedegård and Mads Brandbyge},
journal={Beilstein J. Nanotechnol.},
year={2016},
title={Current-induced runaway vibrations in dehydrogenated
graphene nanoribbons},
volume={7},
pages={68-74},
issn={2190-4286},
doi={10.3762/bjnano.7.8},
publisher={Beilstein-Institut},
URL={https://doi.org/10.3762/bjnano.7.8}
}

@article{Schinabeck2018,
  title = {Hierarchical quantum master equation approach to electronic-vibrational coupling in nonequilibrium transport through nanosystems: Reservoir formulation and application to vibrational instabilities},
  author = {Schinabeck, C. and H\"artle, R. and Thoss, M.},
  journal = {Phys. Rev. B},
  volume = {97},
  issue = {23},
  pages = {235429},
  numpages = {21},
  year = {2018},
  publisher = {American Physical Society},
  doi = {10.1103/PhysRevB.97.235429},
  url = {https://link.aps.org/doi/10.1103/PhysRevB.97.235429}
}

@article{Gunst2013,
  title = {Phonon excitation and instabilities in biased graphene nanoconstrictions},
  author = {Gunst, Tue and L\"u, Jing-Tao and Hedeg\aa{}rd, Per and Brandbyge, Mads},
  journal = {Phys. Rev. B},
  volume = {88},
  issue = {16},
  pages = {161401},
  numpages = {5},
  year = {2013},
  month = {Oct},
  publisher = {American Physical Society},
  doi = {10.1103/PhysRevB.88.161401},
  url = {https://link.aps.org/doi/10.1103/PhysRevB.88.161401}
}

@article{Pecchia2007,
  title = {Theory of heat dissipation in molecular electronics},
  author = {Pecchia, A. and Romano, G. and Di Carlo, A.},
  journal = {Phys. Rev. B},
  volume = {75},
  issue = {3},
  pages = {035401},
  numpages = {10},
  year = {2007},
  month = {Jan},
  publisher = {American Physical Society},
  doi = {10.1103/PhysRevB.75.035401},
  url = {https://link.aps.org/doi/10.1103/PhysRevB.75.035401}
}

@article{Montgomery2002,
  title = {Power dissipation in nanoscale conductors},
  volume = {14},
  ISSN = {0953-8984},
  url = {http://dx.doi.org/10.1088/0953-8984/14/21/312},
  DOI = {10.1088/0953-8984/14/21/312},
  number = {21},
  journal = {J. Phys. Condens. Matter},
  publisher = {IOP Publishing},
  author = {Montgomery,  M J and Todorov,  T N and Sutton,  A P},
  year = {2002},
  month = may,
  pages = {5377–5389}
}

@Article{Ioffe2008,
author={Ioffe, Zvi
and Shamai, Tamar
and Ophir, Ayelet
and Noy, Gilad
and Yutsis, Ilan
and Kfir, Kobi
and Cheshnovsky, Ori
and Selzer, Yoram},
title={Detection of heating in current-carrying molecular junctions by Raman scattering},
journal={Nat. Nanotech.},
year={2008},
month={Dec},
day={01},
volume={3},
number={12},
pages={727-732},
issn={1748-3395},
doi={10.1038/nnano.2008.304},
url={https://doi.org/10.1038/nnano.2008.304}
}

@article{PhysRevB.78.045434,
  title = {Nonequilibrium phonon effects on transport properties through atomic and molecular bridge junctions},
  author = {Asai, Yoshihiro},
  journal = {Phys. Rev. B},
  volume = {78},
  issue = {4},
  pages = {045434},
  numpages = {24},
  year = {2008},
  month = {Jul},
  publisher = {American Physical Society},
  doi = {10.1103/PhysRevB.78.045434},
  url = {https://link.aps.org/doi/10.1103/PhysRevB.78.045434}
}

@article{doi:10.1021/nl801669e,
author = {Tsutsui, Makusu and Taniguchi, Masateru and Kawai, Tomoji},
title = {Local Heating in Metal-Molecule-Metal Junctions},
journal = {Nano Letters},
volume = {8},
number = {10},
pages = {3293-3297},
year = {2008},
doi = {10.1021/nl801669e},
  

URL = { 
    
        https://doi.org/10.1021/nl801669e
    
    

},
eprint = { 
    
        https://doi.org/10.1021/nl801669e
    
    

}}

@article{
doi:10.1126/science.1146556,
author = {Michael Galperin  and Mark A. Ratner  and Abraham Nitzan  and Alessandro Troisi },
title = {Nuclear Coupling and Polarization in Molecular Transport Junctions: Beyond Tunneling to Function},
journal = {Science},
volume = {319},
number = {5866},
pages = {1056-1060},
year = {2008},
doi = {10.1126/science.1146556},
URL = {https://www.science.org/doi/abs/10.1126/science.1146556},
eprint = {https://www.science.org/doi/pdf/10.1126/science.1146556}}

@article{Tao2006,
  title = {Electron transport in molecular junctions},
  volume = {1},
  ISSN = {1748-3395},
  url = {http://dx.doi.org/10.1038/nnano.2006.130},
  DOI = {10.1038/nnano.2006.130},
  number = {3},
  journal = {Nature Nanotechnology},
  publisher = {Springer Science and Business Media LLC},
  author = {Tao,  N. J.},
  year = {2006},
  month = dec,
  pages = {173–181}
}

@article{PhysRevLett.114.096801,
  title = {Current-Induced Forces and Hot Spots in Biased Nanojunctions},
  author = {L\"u, Jing-Tao and Christensen, Rasmus B. and Wang, Jian-Sheng and Hedeg\aa{}rd, Per and Brandbyge, Mads},
  journal = {Phys. Rev. Lett.},
  volume = {114},
  issue = {9},
  pages = {096801},
  numpages = {5},
  year = {2015},
  month = {Mar},
  publisher = {American Physical Society},
  doi = {10.1103/PhysRevLett.114.096801},
  url = {https://link.aps.org/doi/10.1103/PhysRevLett.114.096801}
}

@article{10.1063/5.0019178,
    author = {Preston, Riley J. and Honeychurch, Thomas D. and Kosov, Daniel S.},
    title = {Cooling molecular electronic junctions by AC current},
    journal = {J. Chem. Phys.},
    volume = {153},
    number = {12},
    pages = {121102},
    year = {2020},
    month = {09},
    issn = {0021-9606},
    doi = {10.1063/5.0019178},
    url = {https://doi.org/10.1063/5.0019178},
    eprint = {https://pubs.aip.org/aip/jcp/article-pdf/doi/10.1063/5.0019178/14752835/121102\_1\_online.pdf},
}

@Article{C1CP21161G,
author ="Volkovich, Roie and Härtle, Rainer and Thoss, Michael and Peskin, Uri",
title  = {Bias-controlled selective excitation of vibrational modes in molecular junctions: a route towards mode-selective chemistry},
journal  ="Phys. Chem. Chem. Phys.",
year  ="2011",
volume  ="13",
issue  ="32",
pages  ="14333-14349",
publisher  ="The Royal Society of Chemistry",
doi  ="10.1039/C1CP21161G",
url  ="http://dx.doi.org/10.1039/C1CP21161G"}

@article{Nitzan2003,
  title = {Electron Transport in Molecular Wire Junctions},
  volume = {300},
  ISSN = {1095-9203},
  url = {http://dx.doi.org/10.1126/science.1081572},
  DOI = {10.1126/science.1081572},
  number = {5624},
  journal = {Science},
  publisher = {American Association for the Advancement of Science (AAAS)},
  author = {Nitzan,  Abraham and Ratner,  Mark A.},
  year = {2003},
  month = may,
  pages = {1384–1389}
}

@article{Nitzan2001,
  title = {ELECTRON TRANSMISSION THROUGH MOLECULES AND MOLECULAR INTERFACES},
  volume = {52},
  ISSN = {1545-1593},
  url = {http://dx.doi.org/10.1146/annurev.physchem.52.1.681},
  DOI = {10.1146/annurev.physchem.52.1.681},
  number = {1},
  journal = {Annu. Rev. Phys. Chem.},
  publisher = {Annual Reviews},
  author = {Nitzan,  Abraham},
  year = {2001},
  month = oct,
  pages = {681–750}
}

@article{
doi:10.1126/science.272.5260.385,
author = {R. Martel  and Ph. Avouris  and I.-W. Lyo },
title = {Molecularly Adsorbed Oxygen Species on Si(111)-(7×7): STM-Induced Dissociative Attachment Studies},
journal = {Science},
volume = {272},
number = {5260},
pages = {385-388},
year = {1996},
doi = {10.1126/science.272.5260.385},
URL = {https://www.science.org/doi/abs/10.1126/science.272.5260.385},
eprint = {https://www.science.org/doi/pdf/10.1126/science.272.5260.385}}

@article{PhysRevLett.107.036804,
  title = {Scattering Theory of Current-Induced Forces in Mesoscopic Systems},
  author = {Bode, Niels and Kusminskiy, Silvia Viola and Egger, Reinhold and von Oppen, Felix},
  journal = {Phys. Rev. Lett.},
  volume = {107},
  issue = {3},
  pages = {036804},
  numpages = {4},
  year = {2011},
  month = {Jul},
  publisher = {American Physical Society},
  doi = {10.1103/PhysRevLett.107.036804},
  url = {https://link.aps.org/doi/10.1103/PhysRevLett.107.036804}
}

@incollection{SORBELLO1998159,
title = {Theory of Electromigration},
editor = {Henry Ehrenreich and Frans Spaepen},
series = {Solid State Physics},
publisher = {Academic Press},
volume = {51},
pages = {159-231},
year = {1998},
booktitle = {Solid State Physics},
issn = {0081-1947},
doi = {https://doi.org/10.1016/S0081-1947(08)60191-5},
url = {https://www.sciencedirect.com/science/article/pii/S0081194708601915},
author = {RICHARD S. SORBELLO}
}

@article{Dundas2009,
  title = {Current-driven atomic waterwheels},
  volume = {4},
  ISSN = {1748-3395},
  url = {http://dx.doi.org/10.1038/nnano.2008.411},
  DOI = {10.1038/nnano.2008.411},
  number = {2},
  journal = {	Nat. Nanotechnol.},
  publisher = {Springer Science and Business Media LLC},
  author = {Dundas,  Daniel and McEniry,  Eunan J. and Todorov,  Tchavdar N.},
  year = {2009},
  month = feb,
  pages = {99–102}
}

@article{Huang2007,
  title = {Local ionic and electron heating in single-molecule junctions},
  volume = {2},
  ISSN = {1748-3395},
  url = {http://dx.doi.org/10.1038/nnano.2007.345},
  DOI = {10.1038/nnano.2007.345},
  number = {11},
  journal = {	Nat. Nanotechnol.},
  publisher = {Springer Science and Business Media LLC},
  author = {Huang,  Zhifeng and Chen,  Fang and D’agosta,  Roberto and Bennett,  Peter A. and Di Ventra,  Massimiliano and Tao,  Nongjian},
  year = {2007},
  month = oct,
  pages = {698–703}
}

@article{Segal2002,
  title = {Heating in current carrying molecular junctions},
  volume = {117},
  ISSN = {1089-7690},
  url = {http://dx.doi.org/10.1063/1.1495845},
  DOI = {10.1063/1.1495845},
  number = {8},
  journal = {J. Chem. Phys.},
  publisher = {AIP Publishing},
  author = {Segal,  Dvira and Nitzan,  Abraham},
  year = {2002},
  month = aug,
  pages = {3915–3927}
}

@article{10.1063/5.0222076,
    author = {Mäck, Martin and Thoss, Michael and Rudge, Samuel L.},
    title = {Nonadiabatic dynamics of molecules interacting with metal surfaces: Extending the hierarchical equations of motion and Langevin dynamics approach to position-dependent metal-molecule couplings},
    journal = {J. Chem. Phys.},
    volume = {161},
    number = {6},
    pages = {064106},
    year = {2024},
    month = {08},
    issn = {0021-9606},
    doi = {10.1063/5.0222076},
    url = {https://doi.org/10.1063/5.0222076},
    eprint = {https://pubs.aip.org/aip/jcp/article-pdf/doi/10.1063/5.0222076/20106377/064106\_1\_5.0222076.pdf},
}

@book{Cuevas2010,
  doi = {10.1142/7434},
  url = {https://doi.org/10.1142/7434},
  year = {2010},
  month = jun,
  publisher = {World Scientific},
  author = {Juan Carlos Cuevas and Elke Scheer},
  title = {Molecular Electronics}
}

@article{Rudge2024,
    author = {Rudge, Samuel L. and Kaspar, Christoph and Grether, Robin L. and Wolf, Steffen and Stock, Gerhard and Thoss, Michael},
    title = {Nonadiabatic dynamics of molecules interacting with metal surfaces: A quantum-classical approach based on Langevin dynamics and the hierarchical equations of motion},
    journal = {J. Chem. Phys.},
    volume = {160},
    number = {18},
    pages = {184106},
    year = {2024},
    month = {05},
    issn = {0021-9606},
    doi = {10.1063/5.0204307},
    url = {https://doi.org/10.1063/5.0204307}
}

@article{Hrtle2008,
title={Multimode vibrational effects in single-molecule conductance: A nonequilibrium Green’s function approach},
  url = {https://doi.org/10.1103/physrevb.77.205314},
  year = {2008},
  month = may,
  publisher = {American Physical Society ({APS})},
  volume = {77},
  number = {20},
  author = {R. H\"{a}rtle and C. Benesch and M. Thoss},
  journal = {Phys. Rev. B}
}

@article{PhysRevB.83.115420,
  title = {Stochastic dynamics for a single vibrational mode in molecular junctions},
  author = {Nocera, A. and Perroni, C. A. and Marigliano Ramaglia, V. and Cataudella, V.},
  journal = {Phys. Rev. B},
  volume = {83},
  issue = {11},
  pages = {115420},
  numpages = {16},
  year = {2011},
  month = {Mar},
  publisher = {American Physical Society},
  url = {https://link.aps.org/doi/10.1103/PhysRevB.83.115420}
}

@article{L2011,
  url = {https://doi.org/10.3762/bjnano.2.90},
title={Current-induced dynamics in carbon atomic contacts},
  year = {2011},
  month = dec,
  publisher = {Beilstein Institut},
  volume = {2},
  pages = {814--823},
  author = {Jing-Tao L\"{u} and Tue Gunst and Per Hedeg{\aa}rd and Mads Brandbyge},
  journal = {Beilstein J. Nanotechnol.}
}

@article{L2012,
  url = {https://doi.org/10.1103/physrevb.85.245444},
title={Current-induced atomic dynamics, instabilities, and Raman signals: Quasiclassical Langevin equation approach},
  year = {2012},
  month = jun,
  publisher = {American Physical Society ({APS})},
  volume = {85},
  number = {24},
  author = {Jing-Tao L\"{u} and Mads Brandbyge and Per Hedeg{\aa}rd and Tchavdar N. Todorov and Daniel Dundas},
  journal = {Phys. Rev. B}
}

@article{Sabater2015,
title={Evidence for non-conservative current-induced forces in the breaking of Au and Pt atomic chains},
  url = {https://doi.org/10.3762/bjnano.6.241},
  year = {2015},
  month = {December},
  publisher = {Beilstein Institut},
  volume = {6},
  pages = {2338--2344},
  author = {Carlos Sabater and Carlos Untiedt and Jan M van Ruitenbeek},
  journal = {Beilstein J. Nanotechnol.}
}

@article{Li2016,
  url = {https://doi.org/10.1021/jacs.6b10700},
title={Mechanism for Si–Si Bond Rupture in Single Molecule Junctions},
  year = {2016},
  month = nov,
  publisher = {American Chemical Society ({ACS})},
  volume = {138},
  number = {49},
  pages = {16159--16164},
  author = {Haixing Li and Nathaniel T. Kim and Timothy A. Su and Michael L. Steigerwald and Colin Nuckolls and Pierre Darancet and James L. Leighton and Latha Venkataraman},
  journal = {J. Am. Chem. Soc. }
}

@article{Capozzi2016,
  url = {https://doi.org/10.1021/acs.nanolett.6b01592},
title={Mapping the Transmission Functions of Single-Molecule Junctions},
  year = {2016},
  month = may,
  publisher = {American Chemical Society ({ACS})},
  volume = {16},
  number = {6},
  pages = {3949--3954},
  author = {Brian Capozzi and Jonathan Z. Low and Jianlong Xia and Zhen-Fei Liu and Jeffrey B. Neaton and Luis M. Campos and Latha Venkataraman},
  journal = {Nano Lett.}
}

@article{Li2015,
  url = {https://doi.org/10.1021/ja512523r},
title={Electric Field Breakdown in Single Molecule Junctions},
  year = {2015},
  month = feb,
  publisher = {American Chemical Society ({ACS})},
  volume = {137},
  number = {15},
  pages = {5028--5033},
  author = {Haixing Li and Timothy A. Su and Vivian Zhang and Michael L. Steigerwald and Colin Nuckolls and Latha Venkataraman},
  journal = {J. Am. Chem. Soc.}
}

@Article{e19120647,
AUTHOR = {Sachs, Matthias and Leimkuhler, Benedict and Danos, Vincent},
TITLE = {Langevin Dynamics with Variable Coefficients and Nonconservative Forces: From Stationary States to Numerical Methods},
JOURNAL = {Entropy},
VOLUME = {19},
YEAR = {2017},
NUMBER = {12},
ARTICLE-NUMBER = {647},
URL = {https://www.mdpi.com/1099-4300/19/12/647},
ISSN = {1099-4300},
}

@article{Dou2016,
  url = {https://doi.org/10.1063/1.4965823},
title={Electronic friction near metal surfaces: A case where molecule-metal couplings depend on nuclear coordinates},
  year = {2016},
  month = nov,
  publisher = {{AIP} Publishing},
  volume = {146},
  number = {9},
  author = {Wenjie Dou and Joseph E. Subotnik},
  journal = {J. Chem. Phys.}
}

@article{L2010,
  year = {2010},
title={Blowing the Fuse: Berry’s Phase and Runaway Vibrations in Molecular Conductors},
  month = apr,
  publisher = {American Chemical Society ({ACS})},
  volume = {10},
  number = {5},
  pages = {1657--1663},
  author = {Jing-Tao L\"{u} and Mads Brandbyge and Per Hedeg{\aa}rd},
  journal = {Nano Lett.},
  URL ={https://doi.org/10.1021/nl904233u},    
}

@article{Rudge2023,
  year = {2023},
title={Current-induced forces in nanosystems: A hierarchical equations of motion approach},
  month = mar,
  publisher = {American Physical Society ({APS})},
  volume = {107},
  number = {11},
  author = {Samuel L. Rudge and Yaling Ke and Michael Thoss},
  journal = {Phys. Rev. B},
  url = {https://link.aps.org/doi/10.1103/PhysRevB.107.115416}
}

@article{Preston2021,
  year = {2021},
title={First-passage time theory of activated rate chemical processes in electronic molecular junctions},
  month = mar,
  publisher = {{AIP} Publishing},
  volume = {154},
  number = {11},
  pages = {114108},
  author = {Riley J. Preston and Maxim F. Gelin and Daniel S. Kosov},
  journal = {J. Chem. Phys.},
    url = {https://doi.org/10.1063/5.0045652},
}

@article{Dou2018,
  url = {https://doi.org/10.1063/1.5035412},
title={Perspective: How to understand electronic friction},
  year = {2018},
  month = jun,
  publisher = {{AIP} Publishing},
  volume = {148},
  number = {23},
  author = {Wenjie Dou and Joseph E. Subotnik},
  journal = {J. Chem. Phys.}
}

@article{Dou2016_2,
title={A many-body states picture of electronic friction: The case of multiple orbitals and multiple electronic states},
  url = {https://doi.org/10.1063/1.4959604},
  year = {2016},
  month = aug,
  publisher = {{AIP} Publishing},
  volume = {145},
  number = {5},
  author = {Wenjie Dou and Joseph E. Subotnik},
  journal = {J. Chem. Phys.}
}

@article{Schinabeck2016,
  year = {2016},
title={Hierarchical quantum master equation approach to electronic-vibrational coupling in nonequilibrium transport through nanosystems},
  month = nov,
  publisher = {American Physical Society ({APS})},
  volume = {94},
  number = {20},
  author = {C. Schinabeck and A. Erpenbeck and R. H\"{a}rtle and M. Thoss},
  journal = {Phys. Rev. B},
  url = {https://link.aps.org/doi/10.1103/PhysRevB.94.201407}
}

@article{Erpenbeck2020,
  url = {https://link.aps.org/doi/10.1103/PhysRevB.102.195421},
  year = {2020},
  month = nov,
  publisher = {American Physical Society ({APS})},
  volume = {102},
  number = {19},
  author = {A. Erpenbeck and Y. Ke and U. Peskin and M. Thoss},
  title = {Current-induced dissociation in molecular junctions beyond the paradigm of vibrational heating: The role of antibonding electronic states},
  journal = {Phys. Rev. B}

}

@article{Schinabeck2020,
  url = {https://doi.org/10.1103/physrevb.101.075422},
title={Hierarchical quantum master equation approach to current fluctuations in nonequilibrium charge transport through nanosystems},
  year = {2020},
  month = feb,
  publisher = {American Physical Society ({APS})},
  volume = {101},
  number = {7},
  author = {C. Schinabeck and M. Thoss},
  journal = {Phys. Rev. B}
}

@article{Hrtle2013,
title={Decoherence and lead-induced interdot coupling in nonequilibrium electron transport through interacting quantum dots: A hierarchical quantum master equation approach},
  url = {https://doi.org/10.1103/physrevb.88.235426},
  year = {2013},
  month = dec,
  publisher = {American Physical Society ({APS})},
  volume = {88},
  number = {23},
  author = {R. H\"{a}rtle and G. Cohen and D. R. Reichman and A. J. Millis},
  journal = {Phys. Rev. B}
}

@article{Tanimura1989,
title={Time Evolution of a Quantum System in Contact with a Nearly Gaussian-Markoffian Noise Bath},
  url = {https://doi.org/10.1143/jpsj.58.101},
  year = {1989},
  month = jan,
  publisher = {Physical Society of Japan},
  volume = {58},
  number = {1},
  pages = {101--114},
  author = {Yoshitaka Tanimura and Ryogo Kubo},
  journal = {J. Phys. Soc. Jpn}
}

@article{vonHippel1956,
  url = {https://doi.org/10.1126/science.123.3191.315},
title={Molecular Engineering},
  year = {1956},
  month = feb,
  publisher = {American Association for the Advancement of Science ({AAAS})},
  volume = {123},
  number = {3191},
  pages = {315--317},
  author = {A. von Hippel},
  journal = {Science}
}

@article{Aviram1974,
title={Molecular rectifiers},
  url = {https://doi.org/10.1016/0009-2614(74)85031-1},
  year = {1974},
  month = nov,
  publisher = {Elsevier {BV}},
  volume = {29},
  number = {2},
  pages = {277--283},
  author = {Arieh Aviram and Mark A. Ratner},
  journal = {Chem. Phys. Lett.}
}

@article{Bergfield2013,
title={Forty years of molecular electronics: Non-equilibrium heat and charge transport at the nanoscale},
  url = {https://doi.org/10.1002/pssb.201350048},
  year = {2013},
  month = oct,
  publisher = {Wiley},
  volume = {250},
  number = {11},
  pages = {2249--2266},
  author = {Justin P. Bergfield and Mark A. Ratner},

  journal = {Phys. Status Solidi (b)}
}

@article{Thoss2018,
title={Perspective: Theory of quantum transport in molecular junctions},
  url = {https://doi.org/10.1063/1.5003306},
  year = {2018},
  month = jan,
  publisher = {{AIP} Publishing},
  volume = {148},
  number = {3},
  author = {Michael Thoss and Ferdinand Evers},
  journal = {J. Chem. Phys.}
}

@article{Zimbovskaya2011,
title={Electron transport through molecular junctions},
  url = {https://doi.org/10.1016/j.physrep.2011.08.002},
  year = {2011},
  month = dec,
  publisher = {Elsevier {BV}},
  volume = {509},
  number = {1},
  pages = {1--87},
  author = {Natalya A. Zimbovskaya and Mark R. Pederson},
  journal = {Phys. Rep.}
}

@article{HeadGordon1995,
title={Molecular dynamics with electronic frictions},
  url = {https://doi.org/10.1063/1.469915},
  year = {1995},
  month = dec,
  publisher = {{AIP} Publishing},
  volume = {103},
  number = {23},
  pages = {10137--10145},
  author = {Martin Head-Gordon and John C. Tully},
  journal = {J. Chem. Phys.}
}

@article{Bode2012,
  url = {https://doi.org/10.3762/bjnano.3.15},
title={Current-induced forces in mesoscopic systems: A scattering-matrix approach},
  year = {2012},
  month = feb,
  publisher = {Beilstein Institut},
  volume = {3},
  pages = {144--162},
  author = {Niels Bode and Silvia Viola Kusminskiy and Reinhold Egger and Felix von Oppen},
  journal = {Beilstein J. Nanotechnol.}
}

@article{Maurer2016,
title={Ab initio tensorial electronic friction for molecules on metal surfaces: Nonadiabatic vibrational relaxation},
  url = {https://doi.org/10.1103/physrevb.94.115432},
  year = {2016},
  month = sep,
  publisher = {American Physical Society ({APS})},
  volume = {94},
  number = {11},
  author = {Reinhard J. Maurer and Mikhail Askerka and Victor S. Batista and John C. Tully},
  journal = {Phys. Rev. B}
}

@article{Dou2017_1,
title={Universality of electronic friction: Equivalence of von Oppen's nonequilibrium Green's function approach and the Head-Gordon–Tully model at equilibrium},
  url = {https://doi.org/10.1103/physrevb.96.104305},
  year = {2017},
  month = sep,
  publisher = {American Physical Society ({APS})},
  volume = {96},
  number = {10},
  author = {Wenjie Dou and Joseph E. Subotnik},
  journal = {Phys. Rev. B}
}

@article{Dou2017,
title={Born-Oppenheimer Dynamics, Electronic Friction, and the Inclusion of Electron-Electron Interactions},
  url = {https://doi.org/10.1103/physrevlett.119.046001},
  year = {2017},
  month = jul,
  publisher = {American Physical Society ({APS})},
  volume = {119},
  number = {4},
  author = {Wenjie Dou and Gaohan Miao and Joseph E. Subotnik},
  journal = {Phys. Rev. Lett.}
}

@article{Chen2018,
  url = {https://doi.org/10.1021/acs.jpca.8b09251},
title={Current-Induced Forces for Nonadiabatic Molecular Dynamics},
  year = {2018},
  month = oct,
  publisher = {American Chemical Society ({ACS})},
  volume = {123},
  number = {3},
  pages = {693--701},
  author = {Feng Chen and Kuniyuki Miwa and Michael Galperin},
  journal = {J. Phys. Chem. A}
}

@article{Chen2019,
   url = {https://doi.org/10.1063/1.5095425},
title={Electronic friction in interacting systems},
  year = {2019},
  month = may,
  publisher = {{AIP} Publishing},
  volume = {150},
  number = {17},
  author = {Feng Chen and Kuniyuki Miwa and Michael Galperin},
  journal = {J. Chem. Phys.}
}

@article{Hrtle2018,
   url = {https://doi.org/10.1103/physrevb.98.081404},
title={Cooling by heating in nonequilibrium nanosystems},
  year = {2018},
  month = aug,
  publisher = {American Physical Society ({APS})},
  volume = {98},
  number = {8},
  author = {R. H\"{a}rtle and C. Schinabeck and M. Kulkarni and D. Gelbwaser-Klimovsky and M. Thoss and U. Peskin},
  journal = {Phys. Rev. B}
}

@article{Preston2022,
title={Emergence of negative viscosities and colored noise under current-driven Ehrenfest molecular dynamics},
    url = {https://doi.org/10.1103/physrevb.106.195406},
  year = {2022},
  month = nov,
  publisher = {American Physical Society ({APS})},
  volume = {106},
  number = {19},
  author = {Riley J. Preston and Thomas D. Honeychurch and Daniel S. Kosov},
  journal = {Phys. Rev. B}
}

@article{Todorov2010,
    url = {https://doi.org/10.1103/physrevb.81.075416},
title={Nonconservative generalized current-induced forces},
  year = {2010},
  month = feb,
  publisher = {American Physical Society ({APS})},
  volume = {81},
  number = {7},
  author = {Tchavdar N. Todorov and Daniel Dundas and Eunan J. McEniry},
  journal = {Phys. Rev. B}
}

@article{Todorov2014,
title={Current-induced forces: a simple derivation},
    url = {https://doi.org/10.1088/0143-0807/35/6/065004},
  year = {2014},
  month = sep,
  publisher = {{IOP} Publishing},
  volume = {35},
  number = {6},
  pages = {065004},
  author = {Tchavdar N Todorov and Daniel Dundas and Jing-Tao L\"{u} and Mads Brandbyge and Per Hedeg{\aa}rd},
  journal = {Eur. Phys. J.}
}

@article{Erpenbeck2018,
title={Current-induced bond rupture in single-molecule junctions},
    url = {https://doi.org/10.1103/physrevb.97.235452},
  year = {2018},
  month = jun,
  publisher = {American Physical Society ({APS})},
  volume = {97},
  number = {23},
  author = {A. Erpenbeck and C. Schinabeck and U. Peskin and M. Thoss},
  journal = {Phys. Rev. B}
}

@article{Ke2021,
   url = {https://doi.org/10.1063/5.0053828},
  year = {2021},
title={Unraveling current-induced dissociation mechanisms in single-molecule junctions},
  month = jun,
  publisher = {{AIP} Publishing},
  volume = {154},
  number = {23},
  author = {Yaling Ke and Andr{\'{e}} Erpenbeck and Uri Peskin and Michael Thoss},
  journal = {J. Chem. Phys.}
}

@article{Ke2023,
title={Current-induced bond rupture in single-molecule junctions: Effects of multiple electronic states and vibrational modes},
    url = {https://doi.org/10.1063/5.0155290},
  year = {2023},
  month = jul,
  publisher = {{AIP} Publishing},
  volume = {159},
  number = {2},
  author = {Yaling Ke and Jan Dvo{\v{r}}{\'{a}}k and Martin {\v{C}}{\'{\i}}{\v{z}}ek and Raffaele Borrelli and Michael Thoss},
  journal = {J. Chem. Phys.}
}

@article{Preston2020,
    url = {https://doi.org/10.1103/physrevb.101.155415},
title={Current-induced atomic motion, structural instabilities, and negative temperatures on molecule-electrode interfaces in electronic junctions},
  year = {2020},
  month = apr,
  publisher = {American Physical Society ({APS})},
  volume = {101},
  number = {15},
  author = {Riley J. Preston and Vincent F. Kershaw and Daniel S. Kosov},
  journal = {Phys. Rev. B}
}

@article{Jin2008,
    url = {https://doi.org/10.1063/1.2938087},
title={Exact dynamics of dissipative electronic systems and quantum transport: Hierarchical equations of motion approach},
  year = {2008},
  month = jun,
  publisher = {{AIP} Publishing},
  volume = {128},
  number = {23},
  author = {Jinshuang Jin and Xiao Zheng and YiJing Yan},
  journal = {J. Chem. Phys.}
}

@article{PhysRevLett.68.2512,
  title = {Landauer formula for the current through an interacting electron region},
  author = {Meir, Yigal and Wingreen, Ned S.},
  journal = {Phys. Rev. Lett.},
  volume = {68},
  issue = {16},
  pages = {2512--2515},
  numpages = {0},
  year = {1992},
  month = {Apr},
  publisher = {American Physical Society},
  doi = {10.1103/PhysRevLett.68.2512},
  url = {https://link.aps.org/doi/10.1103/PhysRevLett.68.2512}
}

\end{document}